\documentclass[18pt]{article}
\usepackage{a4wide}
\usepackage{timet,color}
\usepackage[urlcolor=blue,citecolor=black,linkcolor=black]{hyperref}
\usepackage{setspace}
\usepackage{amsmath}
\usepackage{caption}
\usepackage{subcaption}
\usepackage{graphicx}
\usepackage[round]{natbib}
\usepackage{lineno}
\spacing{1.5}
\usepackage[a4paper]{geometry}
\title{Adaptive sequential Monte Carlo for posterior inference and model selection among complex geological priors}

\author{M. Amaya$^1$, N. Linde$^1$, E. Laloy$^2$ \\ {\small
$^1$ Institute of Earth Sciences, University of Lausanne, Switzerland} \\{\small
$^2$ Engineered and Geosystems Analysis, Institute for Environment,} \\ {\small Health and Safety, Belgian Nuclear Research Centre}}
\date{}

\begin{document}

%\label{firstpage}

\maketitle
\begin{small}
Original publication: {\it Geophys.\ J.\ Int.}, 26 April 2021
\end{small}

\begin{footnotesize}
link: (\url{https://academic.oup.com/gji/advance-article/doi/10.1093/gji/ggab170/6253220?login=true})
\end{footnotesize}
\\

\begin{small}
\end{small}

%\linenumbers
\section{Summary}
Bayesian model selection enables comparison and ranking of conceptual subsurface models described by spatial prior models, according to the support provided by available geophysical data. Deep generative neural networks can efficiently encode such complex spatial priors, thereby, allowing for a strong model dimensionality reduction that comes at the price of enhanced non-linearity. In this setting, we explore a recent adaptive sequential Monte Carlo (ASMC) approach that builds on Annealed Importance Sampling (AIS); a method that provides both the posterior probability density function (PDF) and the evidence (a central quantity for Bayesian model selection) through a particle approximation. Both techniques are well suited to parallel computation and rely on importance sampling over a sequence of intermediate distributions, linking the prior and the posterior PDF. Each subsequent distribution is approximated by updating the particle weights and states, compared with the previous approximation, using a small pre-defined number of Markov chain Monte Carlo (MCMC) proposal steps. Compared with AIS, the ASMC method adaptively tunes the tempering between neighboring distributions and performs resampling of particles when the variance of the particle weights becomes too large. We evaluate ASMC using two different conceptual models and associated synthetic cross-hole ground penetrating radar (GPR) tomography data. For the most challenging test case, we find that the ASMC method is faster and more reliable in locating the posterior PDF than state-of-the-art adaptive MCMC. The evidence estimates are found to be robust with respect to the choice of ASMC algorithmic variables and much less sensitive to the model proposal type than MCMC. The variance of the evidence estimates are best estimated by replication of ASMC runs, while approximations based on single runs provide comparable estimates when using a sufficient number of proposal steps in approximating each intermediate distribution.

\section{Introduction}

Markov chain Monte Carlo (MCMC) methods are, for strongly non-linear inverse problems and a limited computational budget, not always able to locate the posterior probability density function (PDF) of interest or to explore it sufficiently. Parallel tempering \citep{earl2005parallel} is a well-known approach to circumvent such issues and it was popularized in geophysics by  \citet{sambridge2014parallel}. Parallel tempering runs multiple interacting chains targeting a sequence of power posteriors including faster moving chains at higher temperatures (i.e., corresponding to less weight being given to the likelihood function). Such chains help to locate significant modes of the posterior distribution that can, through a swapping mechanism, be explored by the chain targeting the posterior distribution of interest for which the temperature is 1. The resulting increase in the ability to bypass local minima and explore multimodal distributions is offset by the need for many parallel chains and a carefully-tailored temperature sequence to ensure efficient mixing among chains.

\citet{neal2001annealed} introduced the annealed importance sampling (AIS) method, which is also well suited to derive information about the posterior PDF of interest when confronted with highly non-linear or multi-modal inverse problems. AIS is a particle method in which many particles (the evolution of each particle is represented by an individual chain) are evolving in parallel. Particle methods rely on the states and weights of a collection of evolving particles to approximate distributions of interest. This is in contrast to MCMC methods in which all states have the same weight and the distribution of interest is approximated by proposal and acceptance mechanisms ensuring that sampling is proportional to the  posterior probability density. In developing AIS, \citet{neal2001annealed} demonstrates how intermediate results obtained by simulated annealing \citep{kirkpatrick1983optimization}, typically used for global optimisation, can be re-interpreted as a sequence of importance sampling steps from approximations of intermediate posterior PDFs at gradually decreasing temperatures (i.e., annealing), thereby, creating a succession of approximations of intermediate distributions between the prior to the posterior distribution of interest. This method has several attractive properties: (1) it inherits from simulated annealing the ability to bypass problems with local minima by initially allowing large steps and efficient exploration before focusing on a more detailed search in areas of high posterior probability; (2) it is well suited for parallelization; (3) the final states and their associated importance weights approximate the posterior distribution; and (4) it offers directly an approximation of the evidence, the central quantity in Bayesian model selection.

Even if AIS is still widely used, it suffers from two main deficiencies: (1) it is very challenging to pre-define an appropriate annealing sequence (i.e., the sequence of inverse temperatures to which the likelihood function is raised) and (2) the populations of importance weights have increasingly higher variances as the AIS run progresses, thereby, increasing the risk of obtaining poor estimates of the posterior PDF and the evidence. Sequential Monte Carlo (SMC) \citep{doucet2011tutorial} represents a family of particle methods that are widely used in science and engineering, particularly for data assimilation tasks, but their use in geophysics has been limited to date (see review by \citet{linde2017uncertainty}). At the most basic level, SMC relies on importance sampling combined with resampling steps which ensures that the particle approximation of the high-dimensional posterior PDF is of sufficient quality. The resampling step tends to reinitialize particles of low probability by states of higher probability, thereby avoiding that computational time is wasted in regions of low posterior density. \citet{zhou2016toward} proposed an adaptive SMC algorithm (referred to hereafter as ASMC) that addresses the limitations of AIS stated above by adaptively tuning the progression between intermediate distributions and by resampling when the variance of the particle weights becomes too large.

The prior PDF has a strong impact on Bayesian geophysical inversion results \citep{hansen2012inverse} and should reflect the existing geological knowledge at a site (see review by \citet{linde2015geological}). One effective way of encoding prior knowledge in a low-dimensional latent vector of uncorrelated parameters is offered by deep generative neural networks \citep{goodfellow2014generative}. \citet{laloy2017inversion} and \citet{laloy2018training} demonstrated using variational autoencoders \citep{kingma2013auto} and generative adversarial networks (GAN) \citep{goodfellow2014generative}, respectively, that the generated realizations of such networks are of high quality and that inversion can be successfully performed on this latent space. The challenge when working with deep generative neural networks is the highly non-linear transform linking the latent variables to the image representation (i.e., the typically gridded model of physical properties). This non-linearity often leads to poor and unreliable convergence when applying gradient-based optimization methods \citep{laloy2019gradient} and inversion on such latent spaces may challenge state-of-the-art MCMC algorithms \citep{laloy2018training}.

Here, we explore the performance of the ASMC method \citep{zhou2016toward} when used together with deep generative networks to approximate evidences and posterior distributions using geophysical data. As examples, we consider crosshole geophysical ground-penetrating radar (GPR) data and GAN-based priors, which implies highly non-linear and challenging inverse problems. In ASMC, the approximations of intermediate posterior distributions is achieved by successively, at each temperature, performing a small number of Markov steps. As model proposals, we consider both an elaborate proposal scheme influenced by evolutionary algorithms and a basic uncorrelated Gaussian proposal. Through these examples, we demonstrate that the ASMC method is: (1) easy to implement in existing MCMC algorithms; (2) well-suited for parallelization; (3) robust to parameter settings and model proposal schemes; (4) providing posterior approximations that can be superior to those offered by state-of-the-art MCMC; and (5) deriving accurate evidence estimations without strong distributional assumptions.

\section{Method}

In our method description below, we rely largely on the notation of \citet{zhou2016toward} who introduced the ASMC algorithm.

\subsection{Bayesian inference and model comparison}

Bayes’ theorem expresses the posterior PDF of a conceptual model $M_{k}$ with parameters $\mathbf{{\theta}}$, given a set of observations $\mathbf{y}$ as:

\begin{equation}
	{\pi}(\mathbf{{\theta}}|\mathbf{y}, M_{k}) = \frac{\pi(\mathbf{{\theta}}|M_{k}) p(\mathbf{y}|\mathbf{{\theta}},M_{k})}{\pi(\mathbf{y}|M_{k})}.
	\label{bayes}
	\end{equation}
	
All the knowledge about the model parameters that is available before considering the data is encapsulated in the prior PDF $\pi(\mathbf{{\theta}}|M_{k})$. The likelihood function $p(\mathbf{y}|\mathbf{{\theta}},M_{k})$ quantifies how likely it is that a specific model realization gave rise to the observations when considering a prescribed error model. The normalizing constant $\pi(\mathbf{y}|M_{k})$ is referred to as the evidence or the marginal likelihood, and it is a multidimensional integral over the parameter space:

\begin{equation}
\pi(\mathbf{y}|M_{k})=\int \pi(\mathbf{{\theta}}|M_{k}) p(\mathbf{y}|\mathbf{{\theta}},M_{k}) d\mathbf{{\theta}}.
\label{evidence_bayes}
\end{equation}

The evidence quantifies the support provided by the data to the conceptual model under consideration, as formalized by the prior PDF, and can be used to rank different conceptual models.
\citet{schoniger2014model} describe and compare different methods to estimate the evidence and found that numerical approaches generate more reliable estimates than mathematical approximations of equation \ref{evidence_bayes} that yield analytical expressions. Recent studies comparing state-of-the-art approaches to evidence estimation in geophysical and hydrogeological contexts include \citet{brunetti2017bayesian} and \citet{brunetti2019hydrogeological}.

\subsection{Adaptive sequential Monte Carlo (ASMC)}\label{ASMC}

\subsubsection{Importance sampling}\label{imp_samp}

Brute Force Monte Carlo (BFMC), also known as the arithmetic mean approach, evaluates many realizations drawn from the prior and the corresponding evidence estimate is their average likelihood. Unfortunately, BFMC suffers from the curse of dimensionality \citep{curtis2001prior} in that most draws from the prior, when considering a handful or more unknown model parameters and high-quality data, have negligible likelihoods. Consequently, high likelihood regions contributing strongly to the mean are poorly sampled, leading to high-variance evidence estimates and frequent underestimation of evidence values as demonstrated by \citet{brunetti2017bayesian}. Throughout this manuscript, a high-variance estimate refers to that obtained by estimators of a mean quantity (e.g., the mean of the sampled likelihoods) for which repeated estimations lead to widely different estimates.

Compared to BFMC, importance sampling offers lower-variance estimates, whereby Monte Carlo samples are drawn proportionally to a so-called importance distribution $q(\mathbf{{\theta}},M_k)$ \citep{Hammersley1964}. In order to sample regions with a high contribution to the mean, this distribution is chosen to be as close as possible to the target distribution; in this case the posterior PDF. To account for the biased sampling procedure, every sample $\theta^i$ drawn from $q(\mathbf{{\theta}},M_k)$ is associated with an importance weight defined as

\begin{equation}\label{IS}
w^i=\frac{\pi(\mathbf{{\theta^i}}|M_{k}) p(\mathbf{y}|\mathbf{{\theta^i}},M_{k})}{q(\mathbf{{\theta^i}},M_k)},
\end{equation}
that determines the corresponding weight in the mean estimation. Assuming that $q(\mathbf{{\theta}},M_k)\neq0$ whenever $\pi(\mathbf{{\theta}}|M_{k}) p(\mathbf{y}|\mathbf{{\theta}},M_{k})\neq0$, and if the number of draws $N\rightarrow \infty$, then the following approximation holds \citep{neal2001annealed}:

\begin{equation}\label{}
\frac{\sum_{i=1}^N w^{i}}{N} \approx \frac{\int \pi(\mathbf{{\theta}}|M_{k}) p(\mathbf{y}|\mathbf{{\theta}},M_{k})d\mathbf{{\theta}} }{\int q(\mathbf{{\theta}},M_k)d\mathbf{{\theta}}}.
\end{equation}

In the particular case of using the prior as the importance distribution (equivalent to BFMC) and noting that its evidence is equal to one (the integral of the prior PDF is 1), the evidence of $M_{k}$ is approximated by the mean of the $N$ weights:

\begin{equation}\label{evid_is}
    \pi(\mathbf{y}|M_k)=\frac{\int \pi(\mathbf{{\theta}}|M_{k}) p(\mathbf{y}|\mathbf{{\theta}},M_{k})d\mathbf{{\theta}} }{\int \pi(\mathbf{{\theta}},M_k)d\mathbf{{\theta}}}\approx\frac{\sum_{i=1}^N w^{i}}{N}=\frac{\sum_{i=1}^N \frac{\pi(\mathbf{{\theta}}^{i}|M_{k}) p(\mathbf{y}|\mathbf{{\theta}}^i,M_{k})}{\pi(\mathbf{{\theta}}^i|M_{k})}}{N}= \frac{\sum_{i=1}^N p(\mathbf{y}|\mathbf{{\theta}}^{i},M_{k})}{N},
\end{equation}
which reduces to the average of the sampled likelihood as discussed above. The importance distribution strongly influences the accuracy of importance sampling and unreliable high-variance estimates are obtained when the importance distribution is far from the target distribution. Therefore, if the prior PDF is markedly different from the posterior PDF, then the quality of the evidence estimate in equation \ref{evid_is} is low. Below, we explain how to obtain low-variance estimates of evidences by relying on a succession of importance sampling steps with importance distributions that are close to intermediate target distributions known as power posteriors.

\subsubsection{Annealed importance sampling (AIS)}

\begin{figure}
\centering
\includegraphics[scale=0.45]{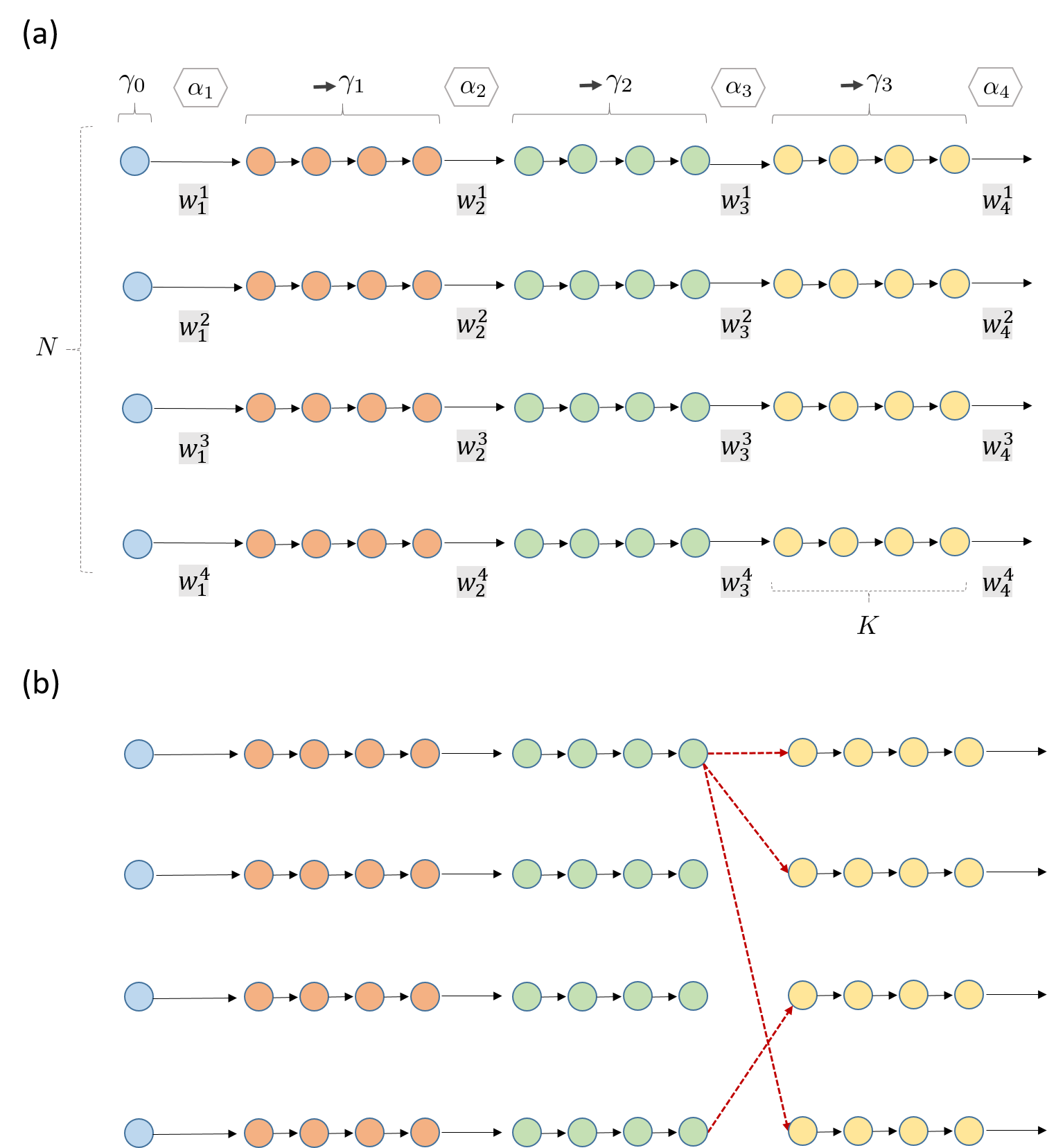}
\caption{(a) Schematic representation of annealed importance sampling (AIS) using $N=4$ particles evolving in parallel. Except for the initialization step, each color represents $K=4$ Markov steps in which the particle system moves from approximating a previous unnormalized power-posterior to a new one. After each $K=4$ Markov steps, the sampled states are used in an importance sampling step to determine the incremental weights $w_t$ associated with the change in the intermediate posterior PDF. (b) In adaptive sequential Monte Carlo (ASMC), one main difference compared with AIS is that the $\alpha$-sequence determining the intermediate posterior distributions is no longer fixed but determined adaptively. Furthermore, resampling occurs when the variance of the weights are too large. Such a resampling step is here visualized with dashed red lines.}
\label{diag}
\end{figure}

Simulated annealing \citep{kirkpatrick1983optimization} is a well-known global optimizer that bypasses local minima by gradually reducing the parameter space exploration using a sequence of intermediate target distributions (i.e., power posteriors characterized by an annealing scheme of successively decreasing temperatures). In developing AIS, \citet{neal2001annealed} took advantage of this sequence of transitional target distributions starting at the prior PDF (infinite temperature) and ending at the posterior PDF (temperature of 1). The algorithm runs in parallel with each chain being interpreted as a particle with an evolving weight and state. From the resulting sequence of intermediate importance weights and states, it is possible to estimate both the posterior PDF and the evidence. AIS shares all the exploratory advantages of simulated annealing and allows for, potentially, high-quality posterior PDF and evidence estimations by creating a smooth path between the prior and the posterior PDF. A schematic visualization of AIS is given in Figure \ref{diag}a. 

In the following, we consider a given conceptual model $M_k$ and suppress the corresponding subindex $k$ for simplicity. The unnormalized power posterior PDFs  $ \left\lbrace\gamma_t(\mathbf{{\theta}}_t|\mathbf{y})\right\rbrace_{t=0}^{T}$ are:

\begin{equation}\label{an_sch}
    {\gamma}_t(\mathbf{{\theta}}_t|\mathbf{y})\equiv\pi(\mathbf{{\theta}}_t)p(\mathbf{y}|\mathbf{{\theta}}_t)^{\alpha_t},
\end{equation}
where $\pi(\mathbf{{\theta}}_t)$ is the prior probability density function and $p(\mathbf{y}|\mathbf{{\theta}}_t)$ the likelihood. The annealing schedule $\alpha_t \in [0,1]$ of inverse temperatures defines these power posteriors, where $\alpha_{t=0}=0$ gives the prior and $\alpha_{t=T}=1$ the posterior PDF. At small $\alpha_t$, the contribution of the likelihood is small and the corresponding power posterior is close to the prior PDF. As $\alpha_t$ grows, the influence of the likelihood on the power posterior increases. We denote $Z_{t}$ as the normalizing constant of the corresponding power posterior, implying that the normalized power PDF is:
\begin{equation}
\pi_{t}(\mathbf{{\theta}}_t|\mathbf{y})=\frac{ \gamma_{t}(\mathbf{{\theta}}_t|\mathbf{y})}{Z_t}.    
\end{equation}

By using $\gamma_{t-1}(\mathbf{{\theta}}_{t-1}|\mathbf{y})$ as an importance distribution for $\gamma_{t}(\mathbf{{\theta}}_{t}|\mathbf{y})$, we define the unnormalized incremental weights $w_t$ for particle $i$ at state $\mathbf{{\theta}}^{i}_{t-1}$ as:

\begin {equation} \label{weig_cont}
    w_t^i=\frac{\gamma_t(\mathbf{{\theta}}^{i}_{t-1}|\mathbf{y})}{\gamma_{t-1}(\mathbf{{\theta}}^{i}_{t-1}|\mathbf{y})}.
\end{equation}
    
Except for the initialization step, the corresponding importance distributions $\gamma_{t-1}(\mathbf{{\theta}}_{t-1}|\mathbf{y})$ are approximated by updating $N$ particles using $K$ Markov steps targeting $\gamma_{t-1}(\mathbf{{\theta}}_{t-1}|\mathbf{y})$ starting at a previous estimation of $\gamma_{t-2}(\mathbf{{\theta}}_{t-2}|\mathbf{y})$. Without these Markov steps, the AIS algorithm would reduce to BFMC. This process is schematized in Figure \ref{diag}a for $N=4$ and $K=4$ .

It is customary to work with normalized weights defined as: 

\begin{equation}\label{norm_weig}
    W_t^i=\frac{W_{t-1}^i {w_t}^i}{\sum_{j=1}^N W_{t-1}^j w_t^j},
\end{equation}
where $W_{t-1}$ are the previously defined normalized weights, that is, $\sum_{i=1}^N W_{t-1}^i = 1$. The final normalized weights $W_T^i$ determine the relative probabilities of each of the final $N$ states, thereby, approximating the posterior distribution through a particle approximation.

\subsubsection{Evidence estimation}    

One major advantage of AIS and ASMC in the context of Bayesian model selection is that the evidence is readily obtained. The ratio of the normalizing constants of two consecutive intermediate distributions $\gamma_t(\mathbf{{\theta}}_t|\mathbf{y})$ and $\gamma_{t-1}(\mathbf{{\theta}}_{t-1}|\mathbf{y})$ is:

\begin {equation}
    \frac{Z_t}{Z_{t-1}} = \frac{\int \gamma_t(\mathbf{{\theta}}_t|\mathbf{y}) d\mathbf{{\theta}}_t}{\int \gamma_{t-1}(\mathbf{{\theta}}_{t-1}|\mathbf{y}) d\mathbf{{\theta}}_{t-1}},
    \end{equation} 
and it can be approximated as \citep{del2006sequential}:

\begin {equation}
    \frac{Z_t}{Z_{t-1}} \approx \sum_{i=1}^N W_{t-1}^{i} w_t^{i}.  
\end{equation}
    
The posterior PDF of interest is the last distribution of the sequence ($\alpha_{t=T}=1$), therefore, its normalizing constant is the evidence, $ Z_T=\pi(\mathbf{y})$. Since the normalizing constant of the prior PDF, $Z_0$, is equal to one, the evidence can be estimated as the product of the intermediate ratios:

\begin {equation}\label{evidence}
    \pi(\mathbf{y})= Z_T = \frac{Z_T}{Z_0} = \prod_{t=1}^{T} \frac{Z_t}{Z_{t-1}}\approx
    \prod_{t=1}^{T} \sum_{i=1}^{N} W_{t-1}^{i}{w_t}^{i}.
\end{equation}

\subsubsection{Adaptive sequence of intermediate distributions}    
\citet{zhou2016toward} introduce several adaptations to AIS leading to the more robust ASMC algorithm that requires much less tuning. The choice of the annealing schedule in equation (\ref{an_sch}) has a strong impact on performance and it is generally difficult to assign a proper $\alpha$-sequence in advance. \citet{zhou2016toward} solve this by introducing an adaptive procedure relying on the conditional effective sample size (CESS):

\begin {equation}\label{cess}
    CESS= N \frac{(\sum_{i=1}^N W_{t-1}^{i}{w_t}^{i})^2}{\sum_{j=1}^N W_{t-1}^{j}({w_t}^{j})^2}.
    \end{equation}

The CESS measures the quality of the current intermediate distribution as an importance distribution to calculate expectations of the following one. To define the next distribution in the sequence (Figure \ref{diag}a), a binary search is performed for the $\alpha$-increment for which the CESS is the closest to a pre-defined target value. The larger this target value is, the better the approximation, but the slower is the algorithm as the $L$ number of intermediate distributions grows. %Note that the selection of the next intermediate distribution only depends on the previous one and its corresponding weights.

\subsubsection{Resampling }

The variance of the importance weights provides an indicator of the quality of the importance estimator. The importance weights invariably diverge over time leading to high variances, for example, because of poor convergence of some particles. To circumvent this, SMC methods rely on resampling \citep{del2006sequential,doucet2011tutorial}. Resampling consists of reinitializing the states of each particle by replicating them according to a probability that is proportional to their current normalized weights. After resampling, the new states are assigned equal weights of $1/N$. Figure \ref{diag}b illustrates a resampling step. The purpose of this operation is to limit the variance of the weights by excluding states with lower weights and replicating those with higher weights. Since high-dimensional posterior distributions are estimated using $N$ particles only, it is essential that all samples contribute meaningfully to this approximation by avoiding regions of very low probability. We rely herein on systematic resampling, which is easy to implement and performs well with respect to alternative resampling schemes \citep{doucet2011tutorial}. The resampling step impacts the variance of estimates \citep{Douc} and it is often beneficial to only perform resampling occasionally. To decide when to apply resampling, we follow standard practice by relying on a quantity that considers the history of the weight variance evolution, namely the effective sample size (ESS) \citep{kong1994sequential}:

\begin {equation}\label{ess}
    ESS_t=\frac{(\sum_{i=1}^N W_{t-1}^{i}w_t^{i})^2}{\sum_{j=1}^N (W_{t-1}^{j})^2(w_t^{j})^2}.
    \end{equation}
    
 The ESS can be interpreted as reflecting the number of effective samples in the particle approximation and resampling is applied when the ESS is lower than a pre-defined threshold.    
 
\subsubsection{Evidence uncertainty estimation}

The most reliable approach to assess uncertainty on evidence estimates is to perform multiple ASMC runs and calculate the resulting variance of the estimates. This is the approach used by \citet{zhou2016toward} when introducing ASMC. Even if such Monte Carlo replication is easily parallelized, it implies a significant computational overhead as the total computational effort grows linearly with the number of replicates. In recent years, progress has been made in obtaining evidence variance estimates from single SMC runs. The first consistent estimator was proposed by \citet{chan2013} and a refined estimator was later introduced by  \citet{lee2018variance}. We consider a modified form of this latter estimator in \citet{doucet2018sequential} that we adopted to account for occasional resampling. The resulting expression should be interpreted as a relative variance contribution of the evidence estimate contribution since the last resampling time:

\begin{equation}\label{var}
\begin{split}
    \frac{V^{N}_{t}}{\left(\eta^N_{t}\right)^2}=\frac{1}{\left(\eta^N_{t}\right)^2} \left(\frac{N}{N-1}\right)^{n} \frac{1}{N(N-1)} \sum_{i=1}^{N} \left[\sum_{j:E_{t}^{j}=i}(NW_{t-1}^j w_{t}^j-\eta^N_{t})\right]^2 ,
\end{split}
\end{equation}
where $\eta^{N}_{t}=\sum_{i=1}^N N W_{t-1}^i w^i_{t}$ and $n$ is the cumulative number of resampling steps that has been performed until \textit{t}. The index $E_t^j$ is the so-called Eve index of particle $j$ at time $t$, which traces the origin of the particles. If no resampling is done, the Eve indices stay constant and are equal to $1:N$. After resampling, the states of the particles are reorganized and the Eve indices change, denoting the original particle that moved to that position. A graphical illustration of this process is given by \citet{lee2018variance}. The number of remaining unique Eve indices along the run can be interpreted as a conservative estimate of the number of independent particles. 

We compute the estimator in equation \ref{var} before each resampling step and at the last step of the ASMC algorithm. We then sum the resulting contributions:

\begin{equation}\label{sigma}
    \sigma_r= \sqrt{\sum_{h=0}^{R}\frac{V^{N}_{h}}{\left(\eta^N_{h}\right)^2}},
\end{equation}
where $R$ is the total number of resampling times. This equation is valid under the assumption that the individual contributions in the sum are independent \citep{Brown}. Hence, we assume here that the particles decorrelate from each other between resampling steps. 

\subsubsection{Markov proposals and acceptance criteria}    

We implemented ASMC within the popular Differential Evolution Adaptive Metropolis ZS (DREAM\textsubscript{(ZS)}) algorithm \citep{laloy2012high}. In this MCMC algorithm, model proposal updates with respect to the present state are drawn proportionally to random differences of past states, thus, helping to better explore the target distribution by automatically determining the scale and direction of the model proposals. If we consider $\mathbf{J}$ as a $m \times d$ dimensional matrix that contains $m$ past states of the chains, where $d$ is the number of parameters, the jump vector for the $i$-th chain is given by \citep{vrugt2016markov}:

\begin{equation}\label{DREAM_prop}
    d\mathbf{{\theta}}^{i}_{A}=\zeta_{d^{*}}+(1_{d^{*}}+\lambda_{d*})\psi(\delta,d^{*})\sum_{j=1}^{\delta} (\mathbf{J}_{A}^{a_{j}}-\mathbf{J}_{A}^{b_{j}}).
\end{equation}

If the current state is $\mathbf{{\theta}}^{i}$, then the candidate point for particle $i$ is $\mathbf{{\theta}}_{prop}^{i}=\mathbf{{\theta}}^{i}+d\mathbf{{\theta}}^{i}$. The number of pairs used to generate the jump is given by $\delta$, and $\mathbf{a}$ and $\mathbf{b}$ are vectors of integers drawn without replacement from $\{1, .., m\}$. The parameters $\zeta$ and $\lambda$ are sampled independently from pre-defined uniform and normal distributions, respectively. This algorithm implements subspace sampling, which implies that only a random subset $A$ of $d^{*}$-dimensions from the original parameter space is updated at each proposal step. The difference between past states is multiplied by a fixed proposal scale referred to as jump rate $\psi(\delta,d^{*})=\frac{2.38}{\sqrt{2\delta d^{*}}} \epsilon$, where $\epsilon$ is an user-defined factor that we introduce to further control the size of the jumps. In contrast to MCMC, ASMC allows straightforward adaptation of the $\epsilon-$factor on-the-go without violating detailed balance condition. This tuning of $\epsilon$ is achieved by using the acceptance rate (AR) of the last $K$ Markov steps to target an acceptance rate above $AR_{min}$. To implement this, $\epsilon$ is initialized to a comparatively large value and a percentage decrease of its value $f$ is made when the acceptance rate falls below $AR_{min}$. For comparison purposes, we also consider standard model proposals given by uncorrelated Gaussian draws centered on the previous state. For this case, the jump vector for the $i$-th chain is given by:

\begin{equation}\label{Gauss_prop}
    d\mathbf{{\theta}}^{i}_{A} \overset{i.i.d.}{\sim} \mathcal{N}_{A}(0,\epsilon^{2}).
\end{equation}

Our considered model proposals are symmetric and the prior PDF is uniform. Consequently, with proper boundary handling, the proposed moves are accepted according to the likelihood ratio \citep{mosegaard1995monte}. The probability to accept each  candidate model during the $K$ Markov steps used to approximate $ \gamma_t(\mathbf{{\theta}}_t|y)$ is:

\begin{equation}\label{Metrop}
    P=min\left\lbrace{1,\frac{p(\mathbf{y}|\mathbf{{\theta}}_{prop})^{\alpha_t}}{p(\mathbf{y}|\mathbf{{\theta}})^{\alpha_t}}}\right\rbrace.
\end{equation}

\subsubsection{Full ASMC algorithm}\label{algor}
The full algorithm is given in Algorithm 1, for which the total number of iterations per considered particle (chain) is equal to $L$ (number of intermediate distributions) $\times$ $K$ (MCMC steps per distribution).

\begin{table*}
    \resizebox{\columnwidth}{!}{
    \begin{tabular}{lll}
    \hline
    \textbf{Algorithm 1: ASCM algorithm adopted from \citet{zhou2016toward}; their algorithm 4. } &       &  \\
    \hline
    Assignment of user-defined variables: &       &  \\
          \multicolumn{2}{l}{\ \ \ \ \ Define number of particles ($N$), optimal CESS ($CESS_{op}$), ESS threshold  ($ESS^{*}$),} \\
          {\ \ \ \ \  number of MCMC iterations at each intermediate distribution ($K$)}, minimal acceptance rate ($AR_{min}$),  &  \\
          {\ \ \ \ \  initial proposal scale factor ($\epsilon$) and its percentage decrease ($f$).} &  \\
    Initialization: Set $t = 0$ &       &  \\
          \multicolumn{2}{l}{\ \ \ \ \ Set $\alpha=0$} &  \\    
          \multicolumn{2}{l}{\ \ \ \ \ Sample $\mathbf{{\theta}}_0$ from the prior $\pi(\mathbf{{\theta}}_t|M_k)$ $N$ times} &  \\
          \multicolumn{2}{l}{\ \ \ \ \ Set the $ N$-dimensional vector of normalized weights $\mathbf{W}_{0}= [\frac{1}{N};\frac{1}{N};...;\frac{1}{N}]$} &  \\        
          \multicolumn{2}{l}{\ \ \ \ \ Set evidence $\pi(\mathbf{y}|M_{k})=1$} &  \\            
    Iteration : Set $t = t+1$ &       &  \\
          \multicolumn{2}{l}{\ \ \ \ \ \textit{Search for incremental distribution }} &  \\
          \multicolumn{3}{l}{\ \ \ \ \ \ \ \ Do binary search for the increment $\Delta\alpha$ that gives the CESS (eq. \ref{cess}) that is the closest to $CESS_{op}$.} \\
          {\ \ \ \ \ \ \ \ Update $\alpha = min(1,\alpha + \Delta\alpha)$ and define the intermediate distribution $\gamma_t(\mathbf{{\theta}}_t|\mathbf{y})=\pi(\mathbf{{\theta}}_t|M_{k})p(\mathbf{y}|\mathbf{{\theta}}_t)^\alpha$}. \\
          \multicolumn{3}{l}{\ \ \ \ \ \ \ \ Compute the weight increments $w_{t}^i$ (eq. \ref{weig_cont}), update and save the normalized weights $W_{t}^i$ (eq. \ref{norm_weig})} \\ %$W_{t}=\frac{W_{t-1} w_{t}}{\sum_{i}^{N} W_{t-1}^{i} w_{t}^{i}}$}\\
          {\ \ \ \ \ \ \ \ and the evidence $\pi(\mathbf{y}|M_{k})=\pi(\mathbf{y}|M_{k}) \sum_{i=1}^{N} W_{t-1}^{i}w_{t}^{i}$ (eq.\ref{evidence}).} \\
          \multicolumn{2}{l}{\ \ \ \ \ \textit{Resampling}} &  \\
          \multicolumn{3}{l}{\ \ \ \ \ \ \ \ Calculate ESS (eq. \ref{ess}), if $ESS < ESS^{*}$ do resampling: re-organize $\mathbf{{\theta}}_t$ states and update $\mathbf{W}_{t}=[\frac{1}{N};\frac{1}{N};...;\frac{1}{N}]$ } \\
          \multicolumn{2}{l}{\ \ \ \ \ \textit{Do $K$ MCMC iterations for each of the $N$ particles (chains)}: }\\
          {\ \ \ \ \ \ \ \  Propose moves $\mathbf{{\theta}}_{prop}$ (eq. \ref{DREAM_prop} and \ref{Gauss_prop}) and accept or reject based on acceptance criterion (eq. \ref{Metrop}) } \\
          {\ \ \ \ \ \ \ \  using $\gamma_t(\mathbf{{\theta}}_t|\mathbf{y})$}.\\
          {\ \ \ \ \ \ \ \  Save the $N$ $\mathbf{{\theta}}$ and their likelihoods.}\\
          {\ \ \ \ \ \ \ \  Set last state as $\mathbf{{\theta}}_{t+1}$}\\
          \multicolumn{2}{l}{\ \ \ \ \ \textit{Tune proposal scale }} &  \\
          \multicolumn{3}{l}{\ \ \ \ \ \ \ \ If acceptance rate $AR < AR_{min}$ then decrease proposal scale factor: $\epsilon=\epsilon*(1-\frac{f}{100})$} \\ 
    Repeat until $\alpha$=1 &       &  \\
\hline
    \end{tabular}    }
  \label{tab:addlabel}%
\end{table*}%

This algorithm has several important strengths: (i) it requires a rather small number of user-defined parameters; (ii) the posterior PDF and the evidence are estimated; (iii) the variance of the weights are used to assess accuracy,  (iv) the adaptation of classical MCMC algorithms into ASMC is straightforward, and (v) the acceptance rate is controlled throughout the inversion.

\subsection{The Laplace-Metropolis method}

MCMC algorithms provide an approximation of the posterior distribution, however, they need to be combined with an additional estimation procedure to provide evidence estimates. For later comparison purposes with ASMC, we mention here the Laplace-Metropolis estimator \citep{lewis1997estimating}, a mathematical approximation of the evidence using a Taylor expansion around the maximum a posteriori (MAP) estimate. Assuming that the posterior PDF is well approximated by a normal distribution, the resulting evidence estimate is:

\begin{equation} \label{LM}
   \pi(\mathbf{y}|M_{k})= (2\pi)^{\frac{d}{2}} |\mathbf{H}(\mathbf{{\theta}}^*)|^{\frac{1}{2}} \pi(\mathbf{{\theta}}^*|M_{k}) p(\mathbf{y}|\mathbf{{\theta}}^*,M_{k}),
\end{equation} 
where $\mathbf{{\theta}}^*$ is the MAP estimate, $d$ is the number of parameters and $|\mathbf{H}(\mathbf{{\theta}}^*)|$ is the determinant of minus the inverse Hessian matrix evaluated at the MAP, which is approximated from the MCMC-based samples from the posterior.

\subsection{From implicit to prescribed geostatistical priors}

Multiple-point statistics (MPS) \citep{mariethoz2014multiple} is a sub-field of geostatistics aiming at producing conditional geostatistical model realizations of high geological realism, thereby, capturing more meaningful connectivity patterns than those offered, for instance, by classical multivariate Gaussian priors \citep{renard2013connectivity}. MPS algorithms produce model realizations that are in agreement with the spatial patterns found in a so-called training image (TI). A TI is a gridded representation of the targeted spatial field obtained from geological information such as outcrops or process-based simulation methods \citep{koltermann1996heterogeneity}. Performing inversion \citep{mariethoz2010bayesian, hansen2012inverse, linde2015geological} and model selection \citep{brunetti2019hydrogeological} based on one or more TIs commonly requires inversion algorithms that work with so-called implicit priors. That is, the MPS algorithm provides model realizations that are drawn proportionally to the prior, but the prior density of a given realization is unknown. Two main issues arise with this approach: (1) the generation of conditional prior realizations may be computationally expensive in MCMC settings when a large number of model proposals are needed, and (2) the implicit prior model precludes the calculation of prior probability densities as needed in many state-of-the-art inversion and model selection methods.

Deep learning \citep{lecun2015deep} applied to geoscientific problems has been growing rapidly in recent years \citep{bergen2019machine,karpatne2018machine}. In particular, deep generative neural networks offer an attractive approach to build an explicit prior PDF from training images \citep{laloy2017inversion,laloy2018training,mosser2017reconstruction,mosser2020stochastic}, that is, a prior for which the prior density of any realization is easily calculated. This is achieved by learning a non-linear transform between a low-dimensional latent space with a prescribed prior (typically an uncorrelated standard normal or bounded uniform prior) and the image space (on which the forward simulations are performed). To do this, the neural network is trained repeatedly with pieces of a large TI or MPS realizations. Such tailor-made model parametrizations achieve significant dimensionality reduction by leveraging spatial patterns in the TI. Inversion is then performed on the latent space and the resulting posterior is mapped, using the trained transform, into a posterior on the original image space (a so-called push-forward operation). We rely on a spatial generative adversarial neural network (SGAN) \citep{jetchev2016texture}, where each dimension of the latent space influences a given region of the generated image space. The network's weights are learned by adversarial training \citep{goodfellow2014generative}. The latter consists of a competition between a so-called discriminator and a generator: the discriminator aims to distinguish fake (i.e., realizations by the generator) and real (i.e., training samples) images, while the generator tries to fool it by generating realizations similar to the training samples. This is mathematically translated in a minimization-maximization problem (see the book by \citet{Goodfellow-et-al-2016}, for details). The main computational effort is related to training and once trained, the computational cost to draw model proposals in the latent space and to map them into the image space (for further forward computations) is very low. The motivation of evaluating ASMC using a deep-learning based parameterization is two-fold: (1) the SGAN parameterization implies strong non-linearity which makes it difficult for MCMC algorithms to converge when performing inversion on the SGAN latent space \citep{laloy2018training}, thus providing challenging test examples for which the added value of ASMC for posterior inference can be demonstrated and (2) to build on recent work \citep{brunetti2019hydrogeological} on MPS-based Bayesian model selection to highlight the value of prescribed priors when performing model selection among MPS-based prior models.

\section{Results}
\subsection{Test examples}

\begin{figure*}
\centering
\includegraphics[scale=0.075]{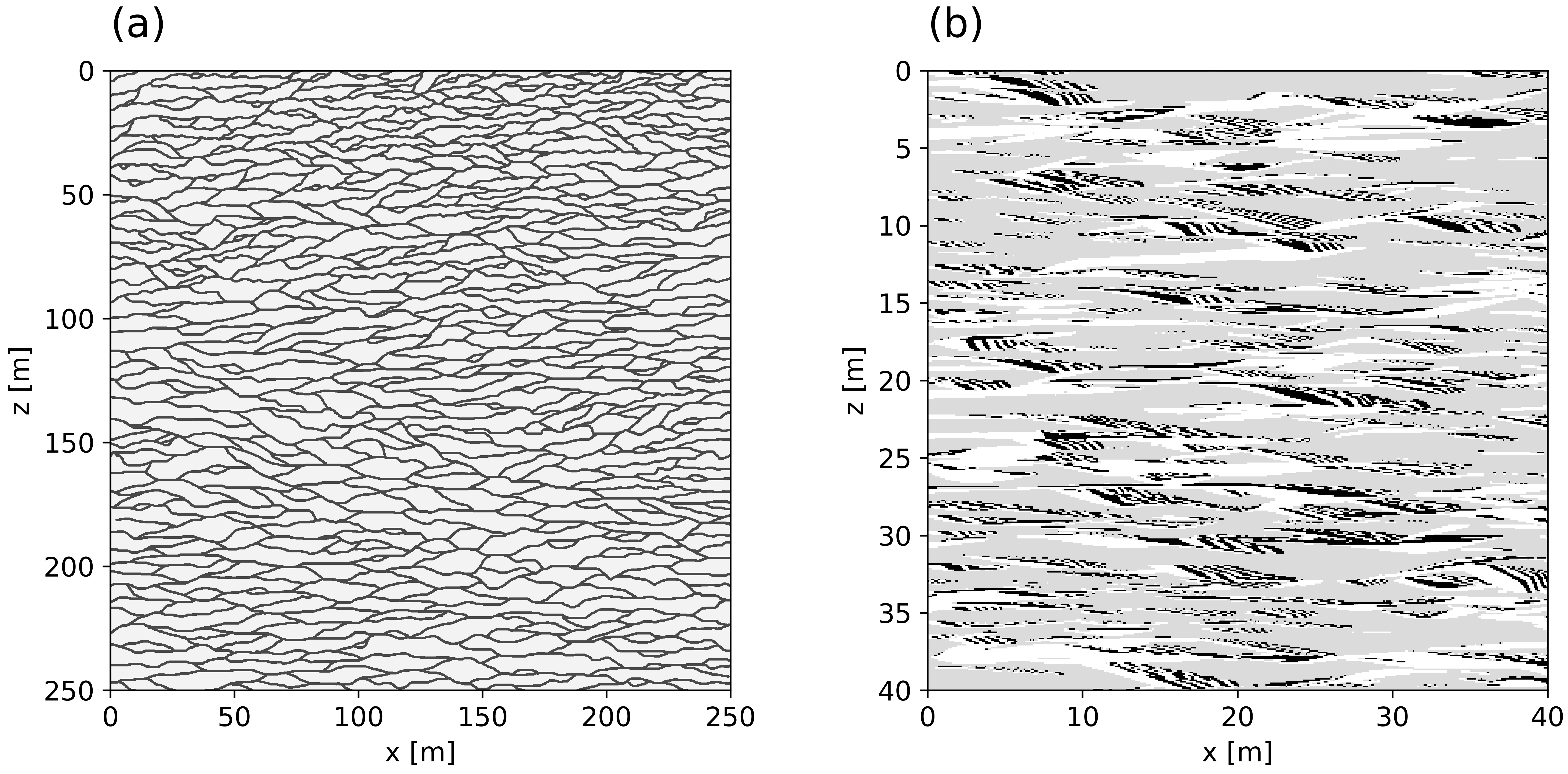}
\caption{Training images: (a) 2500 $\times$ 2500 binary channelized training image (CM1) \citep{zahner2016image} and (b) 400 $\times$ 400 tri-categorical training image representing a braided aquifer (CM2) \citep{pirot2015pseudo}. The discretization of the cells is 0.1 m $\times$ 0.1 m.}
\label{bothTI}
\end{figure*}

\begin{figure}
\centering
\includegraphics[scale=0.8]{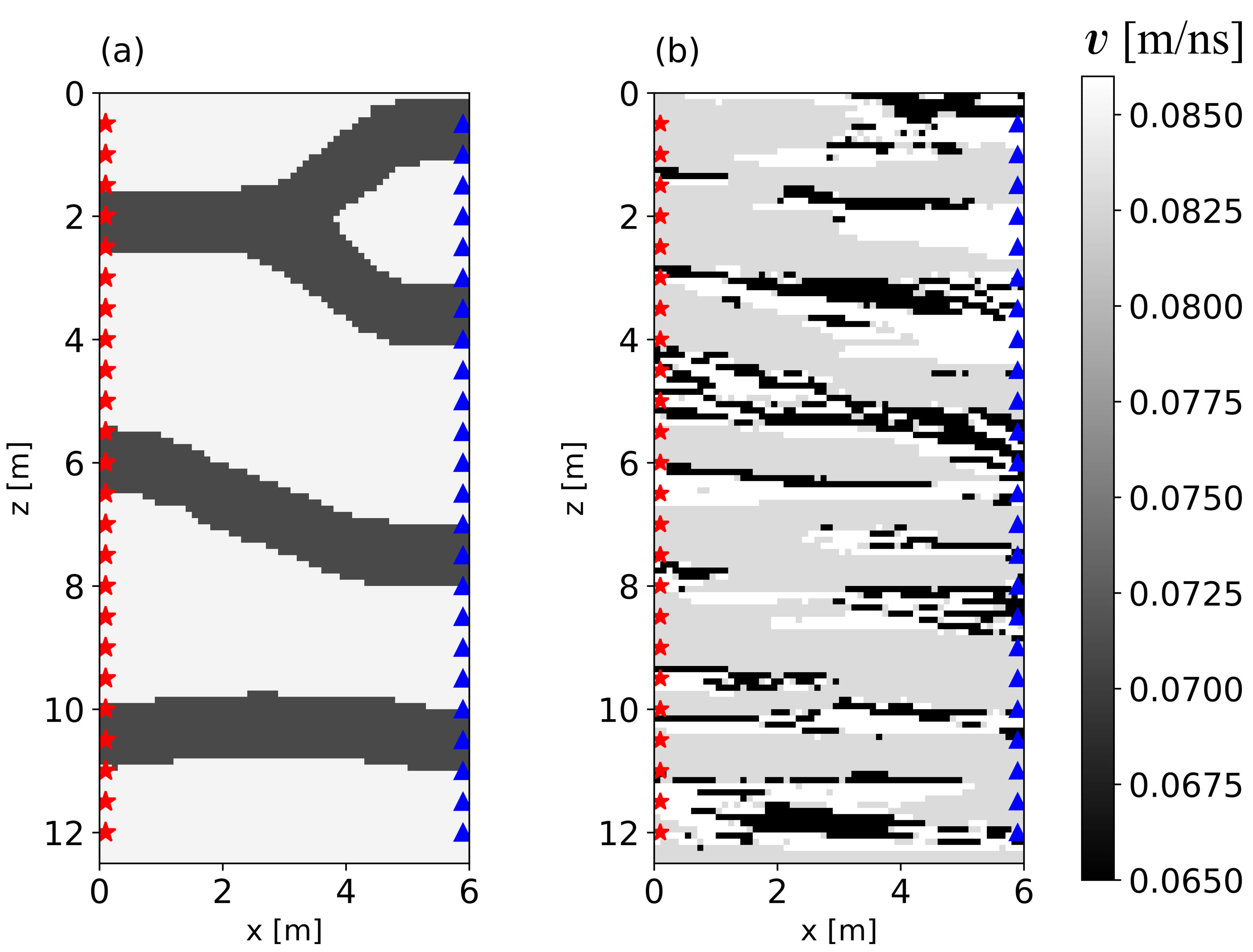}
\caption{Reference models with associated velocities. (a) CM1: channel velocity $v$=0.085 m/ns and matrix velocity $v$=0.071 m/ns. (b) From \citet{pirot2015pseudo} CM2: gray gravel (gray) $v$=0.083 m/ns, open framework (black) $v$=0.065 m/ns and bimodal (white) $v$=0.086 m/ns. Red stars and blue triangles represent GPR sources and receivers, respectively.}
\label{SC}
\end{figure}

Two conceptual 2-D models represented by TIs were used to assess ASMC for inversion and model selection purposes. These TIs are used to train SGANs that generate realizations honoring the multiple-point statistics of the TIs \citep{laloy2018training}. The first conceptual model (Figure \ref{bothTI}a) is represented by a binary channelized training image (CM1) \citep{zahner2016image} and the second one (Figure \ref{bothTI}b) is represented by a tri-categorical training image characterizing braided-river aquifer deposits (CM2) \citep{pirot2015pseudo}. The SGAN generators are assigned uniform priors on the latent space: the CM1-realizations and the more complex CM2-realizations have 15 and 45 latent variables, respectively. All realizations correspond to an image dimension of $129 \times 65$ cells that is cropped to $125 \times 60$, with a discretization of 0.1 m $\times$ 0.1 m (Figure \ref{SC}). 

Our synthetic data correspond to simulated crosshole ground-penetrating radar (GPR) first-arrival travel times with a geometry consisting of two boreholes that are 5.8 m apart. A total of 24 sources and 24 receivers are placed equidistantly every 0.5 meters in depth. First-arrival times were calculated using the \textit{time2d} algorithm by \citet{podvin1991finite}. Following common practice, the data were filtered according to a maximum angle between sources and receivers of 45 degrees \citep{peterson2001pre}, resulting in 444 travel times. In order to assign velocities to each facies, the corresponding dielectric constants were approximated using the complex refractive index method (CRIM) \citep{roth1990calibration}. Representative porosities for CM2 were taken from \citet{pirot2019reduction} and adjusted to CM1 to have the same mean and variance. The two reference models used to produce the synthetic data are shown in Figure \ref{SC}. They were obtained as a randomly chosen realization from the respective SGAN generators. Uncorrelated Gaussian random noise with standard deviation $\sigma=1$ ns was added to the resulting travel times simulated from these models.

\subsection{ASMC performance}\label{asmc_sec}
We first present the parameter settings and the performance of the ASMC algorithm  (section \ref{algor}) using DREAM\textsubscript{(ZS)} proposals (ASMC-DREAM) with $N=40$ particles. To tune the proposal scale, we apply a $20\%$ decrease ($f=0.2$) with $AR_{min}=0.25$. The starting large proposal scale $\epsilon$ is gradually decreased as the annealing progresses (i.e., the inverse temperature $\alpha$ increases towards 1). We implemented adaptive selection of the $\alpha$-sequence, using a binary search defined on a range of $\alpha$-increments from $10^{-5}$ to $10^{-2}$, to find the increments with the $CESS$ that is the closest to the target $CESS_{op}$. The $CESS_{op}/N$ ratio is in practice chosen close to $1$. The closer it is to $1$, the higher the number of intermediate distributions and the larger is the quality of estimates. 
Resampling is applied whenever $ESS/N$ falls below 0.5. Table \ref{tabla1} contains the user-defined parameters and the resulting sequence lengths. The total number of forward simulations of each ASMC run is $N\times K\times L$.  

\begin{figure}
\centering
\includegraphics[scale=0.059]{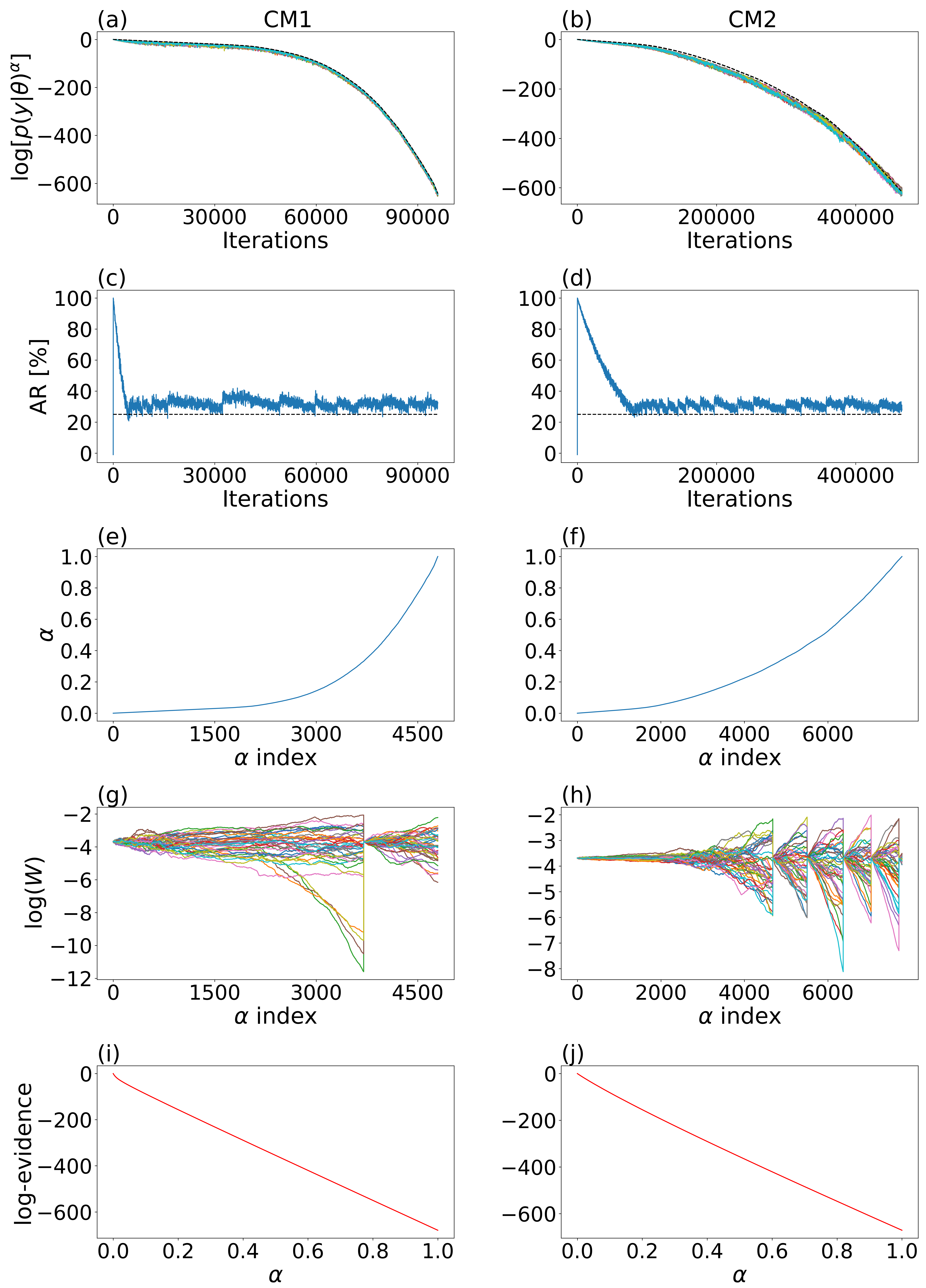}
\caption{Results of ASMC with DREAM\textsubscript{(ZS)} model proposals (ASMC-DREAM) for conceptual models CM1 (left column) and CM2 (right column): (a) and (b) natural logarithm of the likelihood to the power of $\alpha$ vs. iterations per particle. Each color represents a different particle and the black dashed line indicates the logarithm of the likelihood to the power of $\alpha$ calculated using the random noise realization used to noise-contaminate the forward-simulated true model; (c) and (d) acceptance rate vs. iterations per particle, the dashed line indicates a 25$\%$ threshold; (e) and (f) $\alpha$-sequence vs. $\alpha$ index; (h) and (i) natural log-normalized weights vs. $\alpha$ index where each color represents a different particle; (j) and (k) natural log-evidence evolution vs. $\alpha$.
}
\label{ASMC_DREAM}
\end{figure}

Figures \ref{ASMC_DREAM}(a-b) show the evolution of the likelihood raised to the power of the corresponding $\alpha$ in the natural log-scale for CM1 and CM2, respectively. This type of plotting is consistent with the target distribution $\gamma_t(\mathbf{{\theta}}_t|\mathbf{y})$ at each step (equation \ref{an_sch}).  
The black dashed line indicates the target log-likelihood calculated with the random noise realization used to noise-contaminate the forward response of the reference model, raised to the power of the corresponding $\alpha$. Figures \ref{ASMC_DREAM}(c-d) present correspondingly the acceptance rate evolution. As $\alpha$ grows, the acceptance rate for a given jump rate decreases as the targeted posterior distribution gives larger weights to the likelihood. When the acceptance rate falls below $AR_{min}=0.25$, the proposal scale is reduced causing a small increase, after which the acceptance rate starts decreasing again until another reduction of the proposal scale is required, thereby, keeping the acceptance rate in a range between $25\%$ and $40\%$. Figures \ref{ASMC_DREAM}e-f show the optimized sequence of $\alpha$-values defining the intermediate posterior distributions, obtained through a binary search of the $\alpha$-increments. In Figures \ref{ASMC_DREAM}g-h, the logarithm of the normalized weight of each particle is plotted against the $\alpha$-index. Finally, Figures \ref{ASMC_DREAM}i-j shows the evolution of the natural logarithm of the evidence vs. $\alpha$. 

To ensure convergence with the more complex test case CM2, we had to choose a higher $CESS_{op}$ and $K$ than for CM1, which resulted in an approximately 4.7 times longer run. Despite these adaptations, more resampling steps were needed compared to CM1 (see Table \ref{tabla1}), which reinforces the impression that it is a more challenging scenario. The increasing complexity of CM2 is also indicated by the fact that the intermediate target distributions are well-approximated for CM1 (Figure \ref{ASMC_DREAM}a) for which the sampled likelihoods fall close to the dashed line, while this is less the case for CM2 (Figure \ref{ASMC_DREAM}b). However, both test cases reached the target log-likelihood and the resampling fulfills its role of limiting the variance of the weights. 

\begin{figure*}
\centering
\includegraphics[scale=0.059]{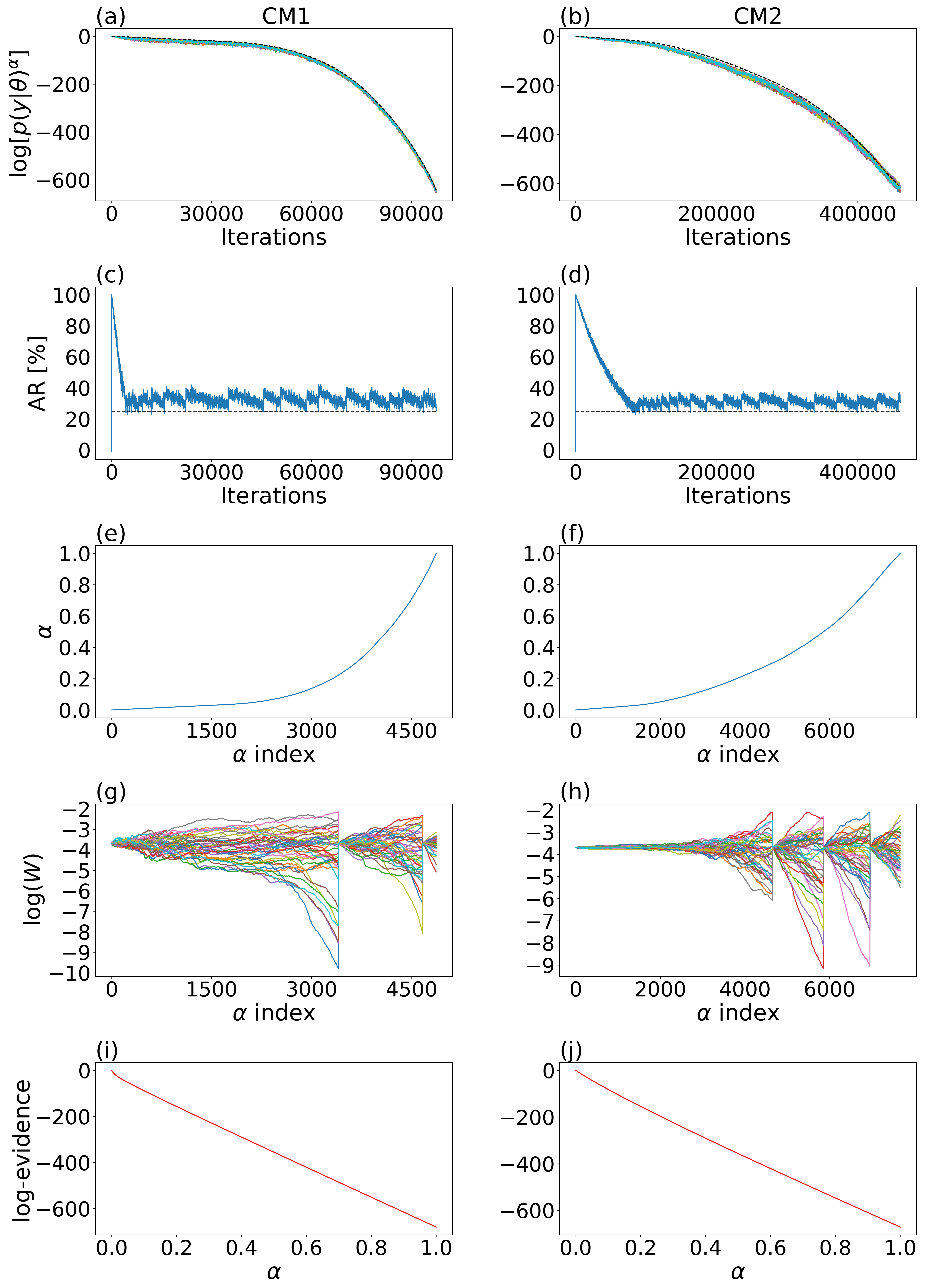}
\caption{Results of ASMC with standard MCMC (ASMC-Gauss) for conceptual models CM1 (left column) and CM2 (right column): (a) and (b) natural logarithm of the likelihood to the power of $\alpha$ vs. iterations per particle. Each color represents a different particle and the black dashed line indicates the logarithm of the likelihood to the power of $\alpha$ calculated using the random noise realization used to noise-contaminate the forward-simulated true model; (c) and (d) acceptance rate vs. iterations per particle, the dashed line indicates a 25$\%$ threshold; (e) and (f) $\alpha$-sequence vs. $\alpha$ index; (h) and (i) natural log-normalized weights vs. $\alpha$ index, each color represents a different particle; (j) and (k) natural log-evidence evolution vs. $\alpha$.
}
\label{ASMC_gauss}
\end{figure*}

Algorithm 1 is applicable to other model proposals than DREAM\textsubscript{(ZS)}. This is demonstrated using standard (vanilla) MCMC model proposals based on uncorrelated random Gaussian perturbations (ASMC-Gauss). In this case, the algorithm starts with a high standard deviation of the centered Gaussian model proposal and it is subsequently decreased when the acceptance rate falls below $25\%$. The user-defined parameters were chosen to be the same as for the ASMC-DREAM tests detailed in Table \ref{tabla1}, leading to a similar sequence length as for ASMC-DREAM. The corresponding results are shown in Figure \ref{ASMC_gauss}. For CM1, ASMC-Gauss needed one more resampling time (Fig. \ref{ASMC_gauss}c) compared to ASMC-DREAM due to a faster increase in the variance of the weights. Otherwise, the performance of ASMC-DREAM (Figure \ref{ASMC_DREAM}) and ASMC-Gauss (Figure \ref{ASMC_gauss}) are very similar. 

\subsection{MCMC performance}
\begin{figure*}
\centering
\includegraphics[scale=0.08]{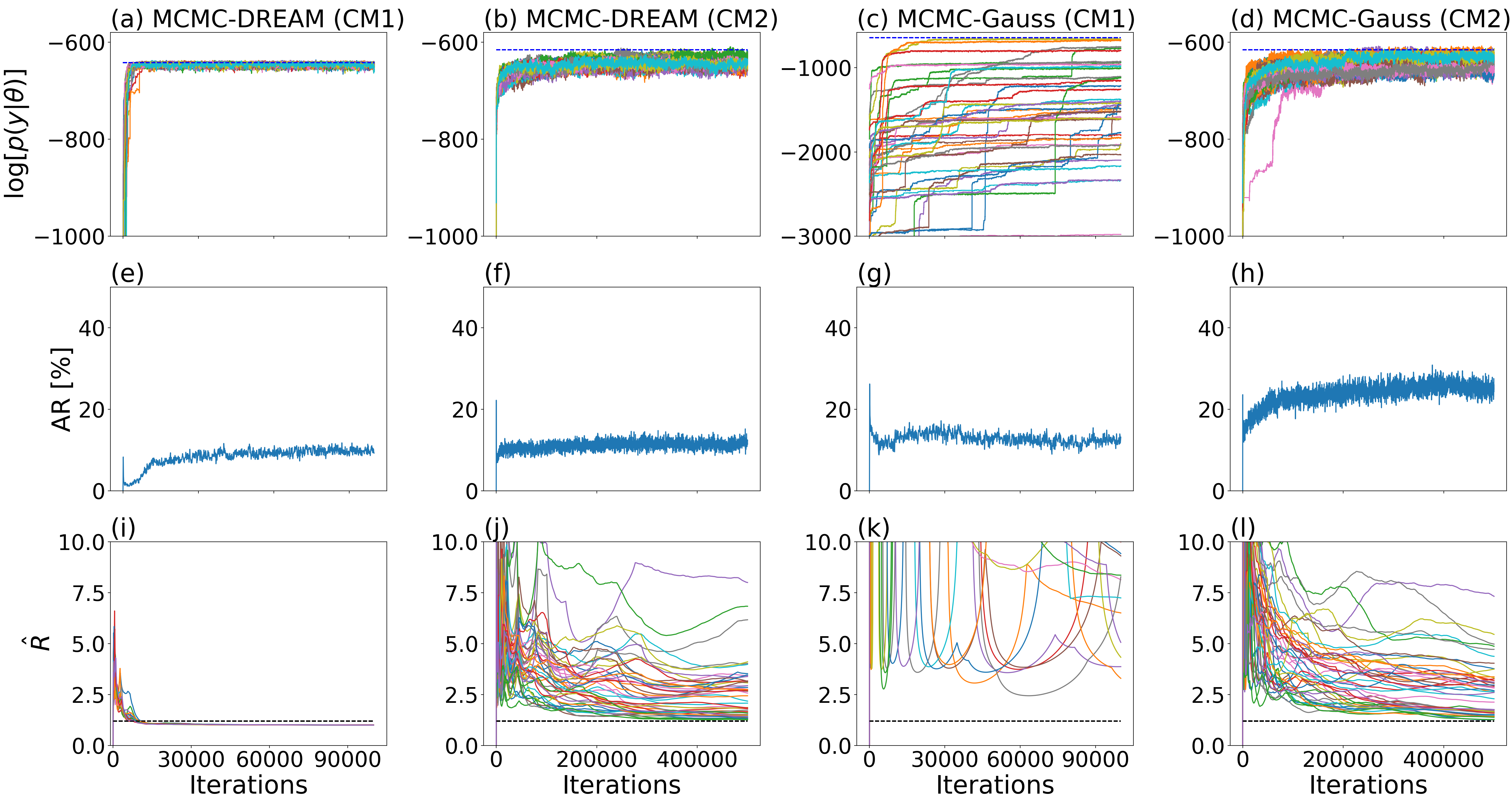}
\caption{MCMC inversion results from DREAM\textsubscript{(ZS)} (MCMC-DREAM) and standard MCMC with Gaussian model proposals (MCMC-Gauss) for conceptual models CM1 and CM2. (a)-(d) the natural logarithm of the likelihood vs. iterations, where each color represents a different particle and the black dashed line indicates the log-likelihood calculated using the random noise realization, (e)-(h) the acceptance rate evolution, and (i)-(l) the evolution of the potential scale reduction factor $\hat{R}$ with each color representing a different parameter and the black dashed lines indicating the value below which convergence is declared ($\hat{R}=1.2$).}
\label{MCMC_without_ASMC}
\end{figure*}

For comparative purposes, we also perform MCMC inversions (no ASMC) using 40 chains and a similar number of forward simulations. Again, we consider two tests: one using DREAM\textsubscript{(ZS)} (MCMC-DREAM) and one with  random Gaussian perturbations (MCMC-Gauss). Extensive manual tuning of the inversion parameters was needed to achieve satisfactory results. Figure \ref{MCMC_without_ASMC} shows the results obtained for conceptual models CM1 and CM2. The log-likelihood evolution is shown in Figures \ref{MCMC_without_ASMC}a-d and the acceptance rate in Figures \ref{MCMC_without_ASMC}e-h. In order to assess convergence, the potential scale reduction factor $\hat{R}$ is calculated \citep{gelman1992inference} and plotted in Figures \ref{MCMC_without_ASMC}i-l, with convergence declared when $\hat{R}$ is below $1.2$ for all model parameters. 

The only MCMC run reaching convergence is MCMC-DREAM for CM1 at around 10,000 iterations. For this conceptual model, the results obtained with MCMC-Gauss are unsatisfactory with only a few of the chains approaching the target likelihood, while the others are trapped in local minima, thereby, demonstrating a vastly superior performance of MCMC-DREAM compared with MCMC-Gauss. For CM2, none of the MCMC inversions converge within the allotted computational time, as $\hat{R}$ does not fall below 1.2. This is also reflected in the likelihood evolution: the majority of sampled likelihoods remains below the target likelihood along the run. To summarize, we find for a similar computational budget that the ASMC algorithm reaches the target likelihood for both conceptual models and model proposal types, while the MCMC runs only approximate the target likelihood for CM1 using MCMC-DREAM.

\begin{table*}
%\begin{minipage}{106mm}
\caption{ASMC user-defined parameters and resulting sequence length for conceptual models CM1 and CM2.}
\label{}
\centering
\begin{tabular}{@{}lcccc}
%\hline
&   ASMC-DREAM & ASMC-DREAM&   ASMC-Gauss & ASMC-Gauss \\%[-5pt]
&  CM1  & CM2 &  CM1  & CM2  \\[2pt]
\hline
Particles ($N$) & 40 & 40 & 40 & 40\\ [2pt]
$CESS_{op}/N$ & 0.999993 & 0.999996 & 0.999993 & 0.999996 \\[2pt]
$ESS^{*}/N$ & 0.5 & 0.5 & 0.5 & 0.5\\[2pt]
$AR_{min}$ & 25\% & 25\% & 25\%  & 25\% \\[2pt]
$K$ iterations  & 20 & 60 & 20 & 60 \\[2pt]
$L$ intermediate distributions & 4798  & 7775 & 4871 & 7673 \\[2pt]
Iterations per particle & 95960 & 466500 & 97420 & 460380 \\[2pt]
Resampling times  & 1 & 5 & 2 & 3 \\[2pt]
Total numerical demand [$\times10^5$] & 38.384 & 186.600 & 38.968 & 184.152 \\[2pt]
\hline
\label{tabla1}
\end{tabular}
%\end{minipage}
\end{table*}

\subsection{Posterior distributions}

We focus now on the posterior approximations obtained with ASMC-DREAM and MCMC-DREAM. For MCMC-DREAM, the posterior is obtained by first removing the so-called burn-in period, that is, the number of iterations needed to reach the target likelihood, from which it starts to sample from the posterior PDF. The remaining samples contribute equally to the posterior estimations. This is not the case for ASMC, for which the posterior PDF is approximated by the last states and weights of the particles (chains). 

\begin{figure*}
\centering
\includegraphics[scale=0.17]{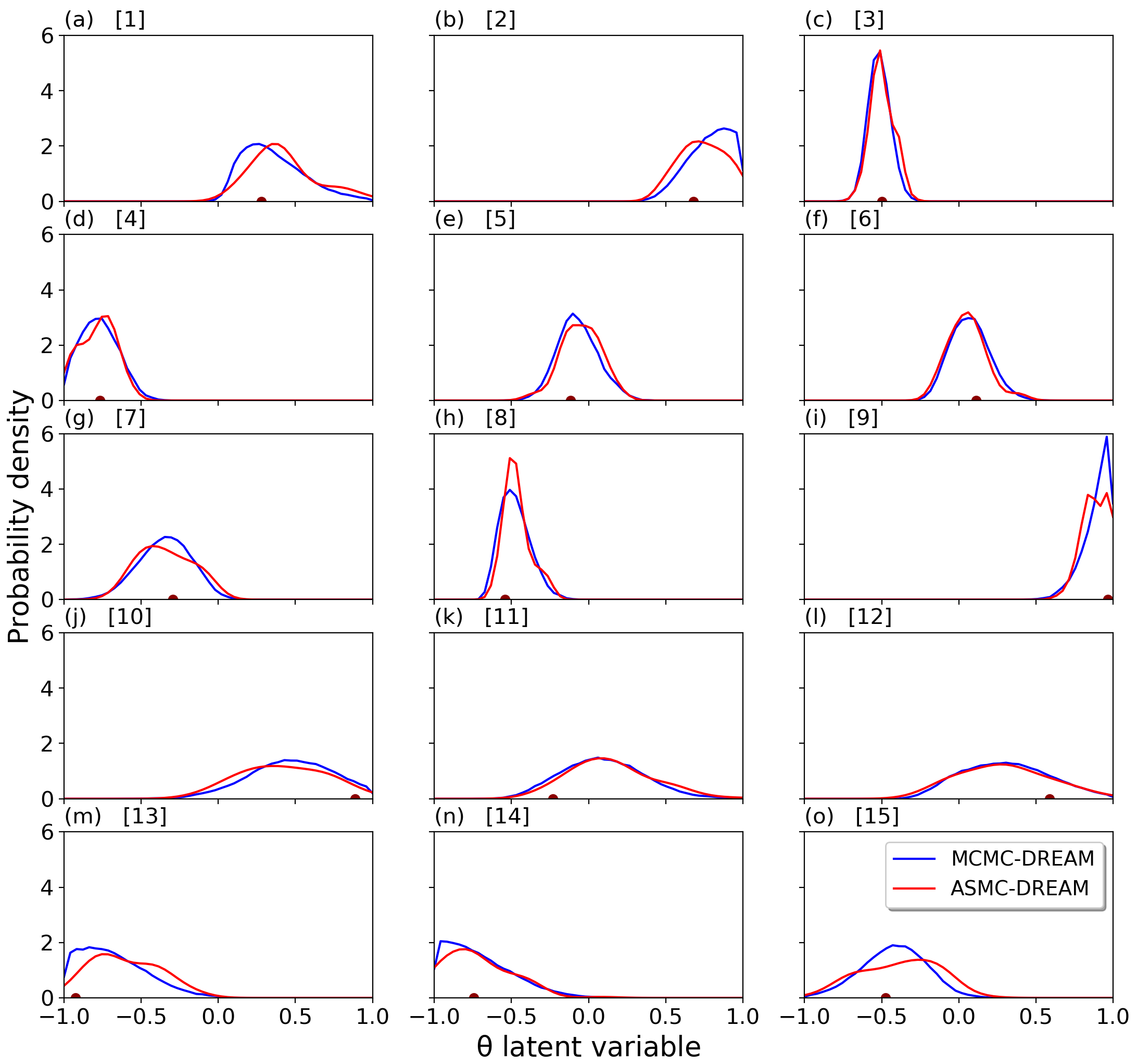}
\caption{Estimated marginal posterior distributions for CM1 using ASMC with DREAM\textsubscript{(ZS)}-proposal (ASMC-DREAM) and regular DREAM\textsubscript{(ZS)} (MCMC-DREAM) with a comparable number of forward computations. Results are shown for all latent model parameters that have bounded uniform priors between -1 and 1.
}
\label{Posterior_CM1}
\end{figure*}

For a smoother representation of the posterior PDF approximated by ASMC, we applied kernel density estimation (KDE) \citep{scott2015multivariate}. Figure \ref{Posterior_CM1} compares the estimated posteriors for CM1. The KDE bandwidth impacts on the level of smoothing, that we chose to kept fixed for the $15$ parameter posteriors. Nevertheless, the estimated posteriors are overall very similar, which suggests that ASMC provides a good estimation of the posterior. No comparison is provided for CM2 as the MCMC-DREAM algorithm did not converge, neither in terms of reaching the target likelihood nor in terms of exploration of the posterior PDF.

\begin{figure*}
\centering
\includegraphics[scale=0.05]{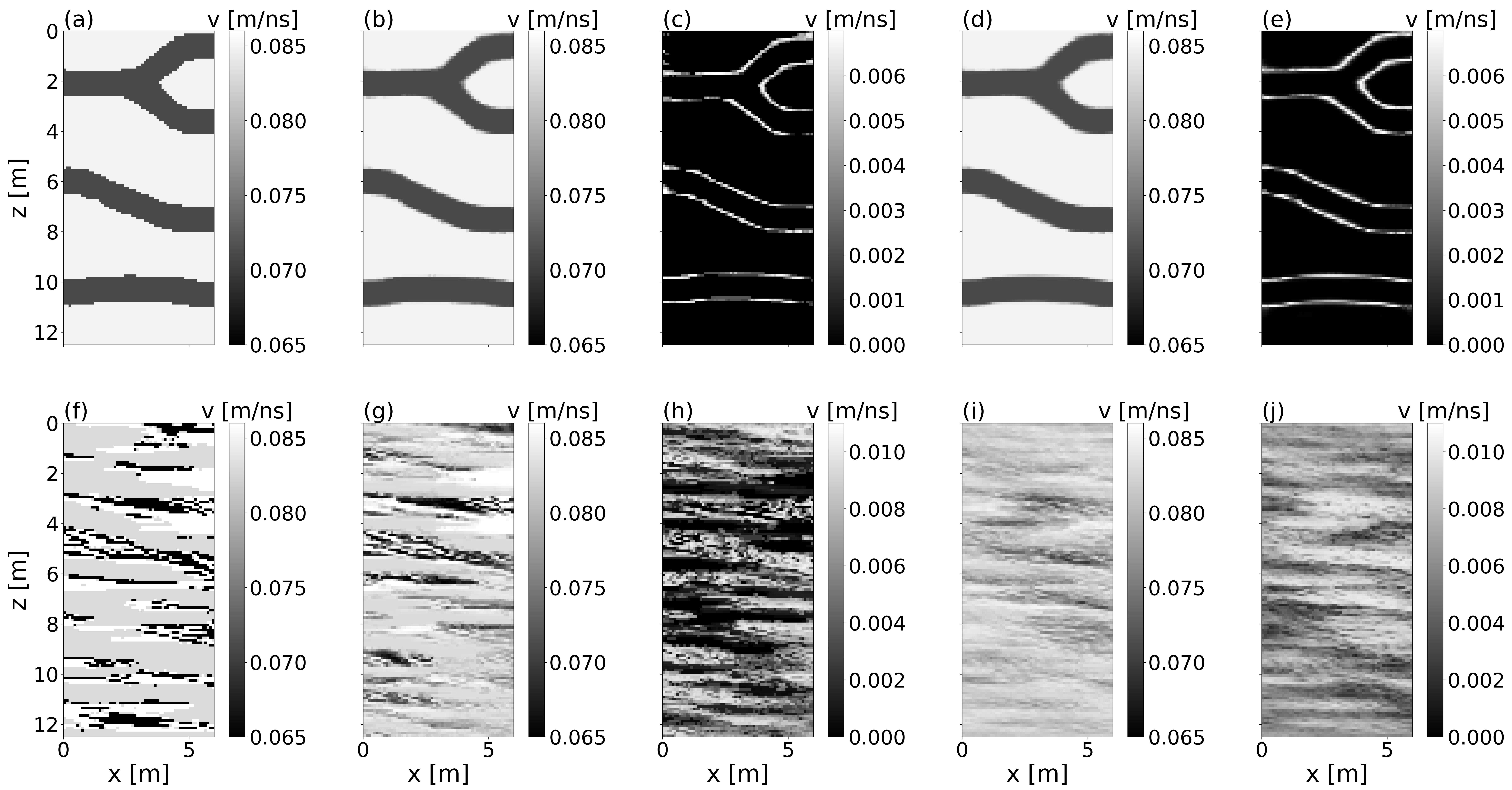}
\caption{Reference model for (a) CM1 and (f) CM2; mean of the weighted final states from ASMC-DREAM for (b) CM1 and (g) CM2; standard deviations of the corresponding weighted final states for (c) CM1 and (h) CM2; mean of the second half of the MCMC chains obtained with MCMC-DREAM  for (d) CM1 and (i) CM2 (not converged); corresponding standard deviations for (e) CM1 and (j) CM2 (not converged).  
}
\label{last_models}
\end{figure*}

We now consider the posterior means and variances in the image space by translating the posterior realizations in the latent space using the SGAN generator. For ASMC-DREAM, the mean and standard deviation images correspond to the last states of the chains weighted by their weights. For MCMC-DREAM, the mean and standard deviation images are obtained using the equally weighted states in the second half of the chains. The means and standard deviations for CM1 are very similar for ASMC-DREAM (Figure \ref{last_models}b-c) and MCMC-DREAM (Figure \ref{last_models}d-e) that both approximate the true model very well (Figure \ref{last_models}a). For CM2, we see a much better defined mean model and smaller standard deviations for ASMC-DREAM (Figure \ref{last_models}g-h). The poorer approximations by MCMC-DREAM \ref{last_models}i-h) is a direct consequence of the fact that this run did not converge. Table \ref{lik_evid} shows the log-likelihood range for the different inversions. For MCMC-DREAM, the second halves of the chains are considered for the range, while only the last states of the particles are considered for ASMC-DREAM.

\subsection{Evidence estimation}\label{evid_sect}
Even if the theoretical basis of the ASMC method for evidence estimation is well-established \citep{zhou2016toward}, we start this section by considering a simple example that allows for comparison with BFMC (see section \ref{imp_samp}). We consider CM1 in a high-noise setting using uncorrelated Gaussian random noise with standard deviation $\sigma= 15$ ns. This is certainly an unrealistically high noise level, but it allows us to obtain reliable evidence estimates through BFMC using 2 million prior samples. The resulting log-evidence obtained by BFMC is -1798.92, while the corresponding ASMC-DREAM run using $K=5$ and $CESS_{op}/N=0.9999$ (resulting in 1100 iterations per particle) led to a log-evidence estimate of -1798.86, which is practically identical to the BFMC estimate.

After having established that our ASMC implementation provides accurate evidence estimation by comparison with BFMC, we now return to the original low-noise $\sigma = 1$ ns setting. For the test examples considered in the previous sections, the evidence estimates obtained with ASMC-DREAM and ASMC-Gauss given in Table \ref{lik_evid} (i.e., the last computed values shown in Figures \ref{ASMC_DREAM}i-j and \ref{ASMC_gauss}i-j) are very close to each other. For comparison purposes, we also calculate the Laplace-Metropolis evidence estimator (LM) using the MCMC-DREAM inversion results (equation \ref{LM}). This is done for CM1 only as MCMC-DREAM did not converge for CM2. The Laplace-Metropolis estimate (Table \ref{lik_evid}) is only slightly lower than the ASMC-DREAM and ASMC-Gauss estimates. The close agreement between ASMC-DREAM and ASMC-Gauss, and the close agreement considering the simplifying assumptions of the Laplace-Metropolis method, suggest again that the results obtained with ASMC are accurate.
 
\begin{figure*}
\centering
\includegraphics[scale=0.12]{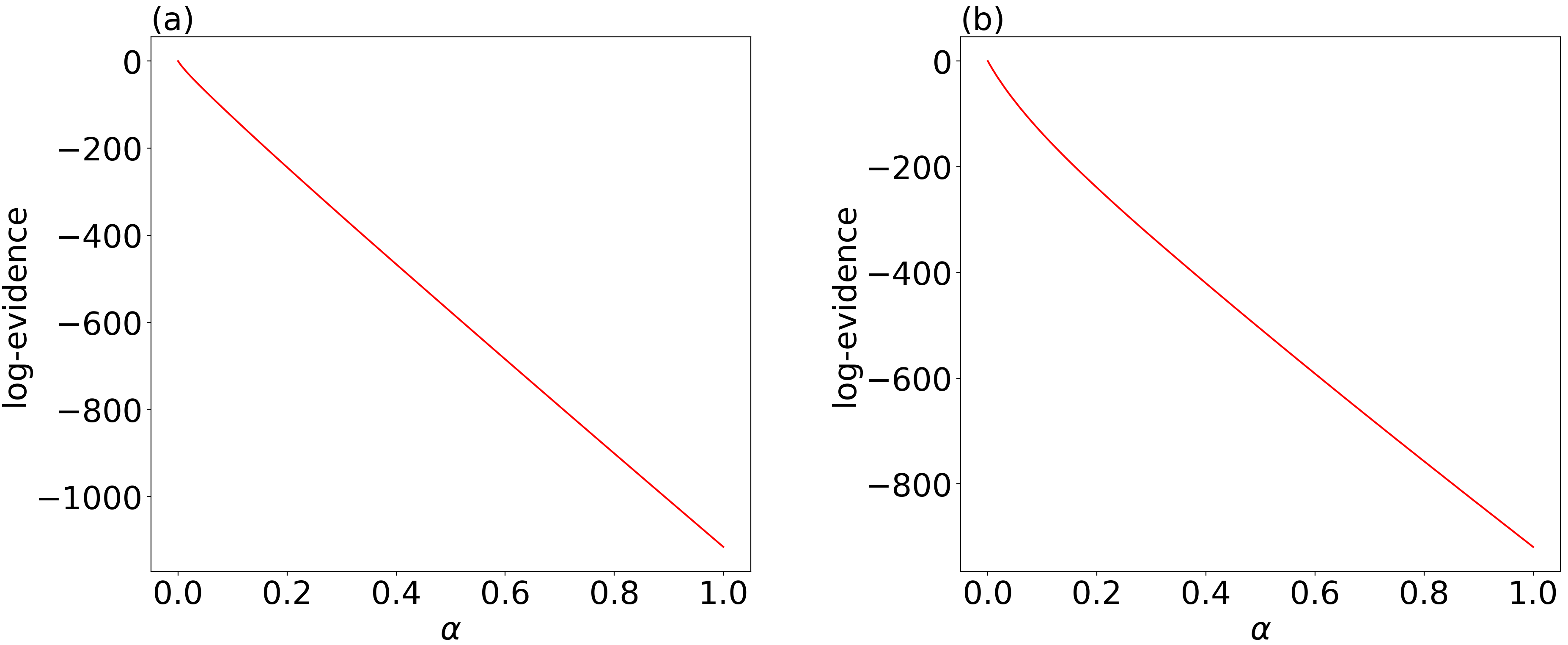}
\caption{ASMC-DREAM evidence evolution with respect to the $\alpha$-sequence evolution when making incorrect assumptions about the underlying conceptual model: (a) CM1-based prior in the inversion using data generated from a prior realization from CM2, and (b) CM2-based prior in the inversion using data generated from a prior realization from CM1. 
}
\label{wrong_assum}
\end{figure*}

 Until now, we have considered that the right conceptual (prior) model was used in the inversions. That is, the noise-contaminated data were generated with a realization of the assumed prior PDF. We now consider how the evidence changes if we make the wrong assumption, that is, use the noise-contaminated data generated from a prior draw of another conceptual model.   
In Figure \ref{wrong_assum} we display the evidence evolution for two such incorrect scenarios using ASMC-DREAM with combinations of CM1 and CM2 in the data generation and inversion process. The resulting log-evidence estimates (Table \ref{lik_evid}) are many hundreds of times smaller than the estimations obtained by making the right assumption, suggesting in these simple scenarios that the true conceptual model can easily be inferred if it is in the set of considered conceptual models.

\begin{table*}
%\begin{minipage}{106mm}
\caption{Natural log-likelihood range,  natural log-evidence estimation and number of resampling steps for the different inversion cases. The log-likelihoods of the reference models are -$642.34$ (CM1) and $-616.00$ (CM2).}
\label{lik_evid}
\centering
\begin{tabular}{@{}lccc}
%\hline
 & Log-likelihood  & Log-evidence  & Resampling \\%[-6pt]
 & range & estimation & times\\
\hline
CM1 inv - CM1 data/ ASMC-DREAM      & [-652.03; -641.02]  & -679.48  & 1 \\[2pt]
CM1 inv - CM1 data/ MCMC-DREAM     & [-666.07; -636.71] & -678.39\small{\textit{(LM)}}    & - \\[2pt]
CM1 inv - CM1 data/ ASMC-Gauss     & [-654.79; -640.65]  & -679.80  & 2\\[2pt]
CM2 inv - CM2 data/ ASMC-DREAM    & [-628.60; -603.91] & -671.18 & 5 \\[2pt]
CM2 inv - CM2 data/ MCMC-DREAM    & [-682.90; -612.23] & - & -\\[2pt]
CM2 inv - CM2 data/ ASMC-Gauss   & [-638.64; -611.15] & -671.49 & 3\\[2pt]
CM1 inv - CM2 data/ ASMC-DREAM    & [-1086.42;-1063.34] & -1115.76 & 5 \\[2pt]
CM2 inv - CM1 data/ ASMC-DREAM    & [-831.70; -795.19] & -919.17 & 9 \\[2pt]
\hline
\end{tabular}
%\end{minipage}
\end{table*}

\subsection{Evidence uncertainty quantification}

\begin{figure*}
    \centering
    \includegraphics[scale=0.04]{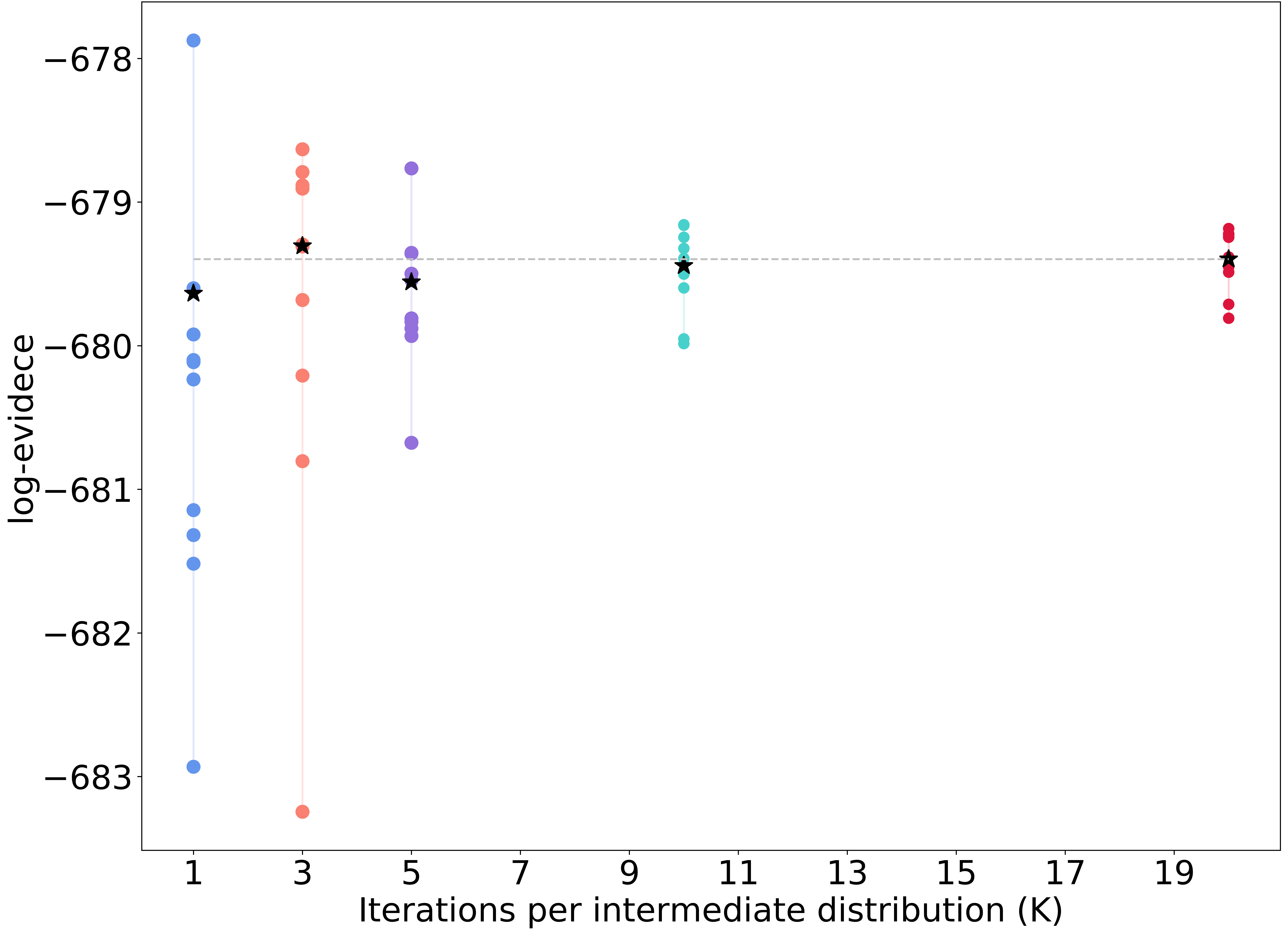}
    \caption{Natural log-evidence estimations for ten replications of the ASMC-DREAM algorithm applied to CM1 using $K=1,3,5,10,20$ iterations per intermediate distribution, where each colored point denotes a given replication. The gray dashed line represents the mean of the $K=20$ replications and the black stars the corresponding mean for each $K$.}
\label{replications}    
\end{figure*}

We first assess the uncertainty of the evidence estimations by performing Monte Carlo replication. For the low noise ASMC-DREAM tests shown in section \ref{asmc_sec}, we performed ten separate runs of ASMC-DREAM for CM1 and five for CM2. We varied $K$ and kept all other parameters fixed. Figure \ref{replications} shows the corresponding evidence estimations for CM1 and their means in logarithmic units. Table \ref{evid_err} shows the relative standard deviation for both conceptual models. For CM1, it decreases almost by a factor of $10$ when moving from $K=1$ to $K=20$. For this case, even $K=1$ leads to rather high-quality estimates with a relative standard deviation of 1.72. The decrease is less abrupt for CM2 when increasing $K=5$ to $K=60$.

\begin{figure*}
    \centering 
    \includegraphics[scale=0.045]{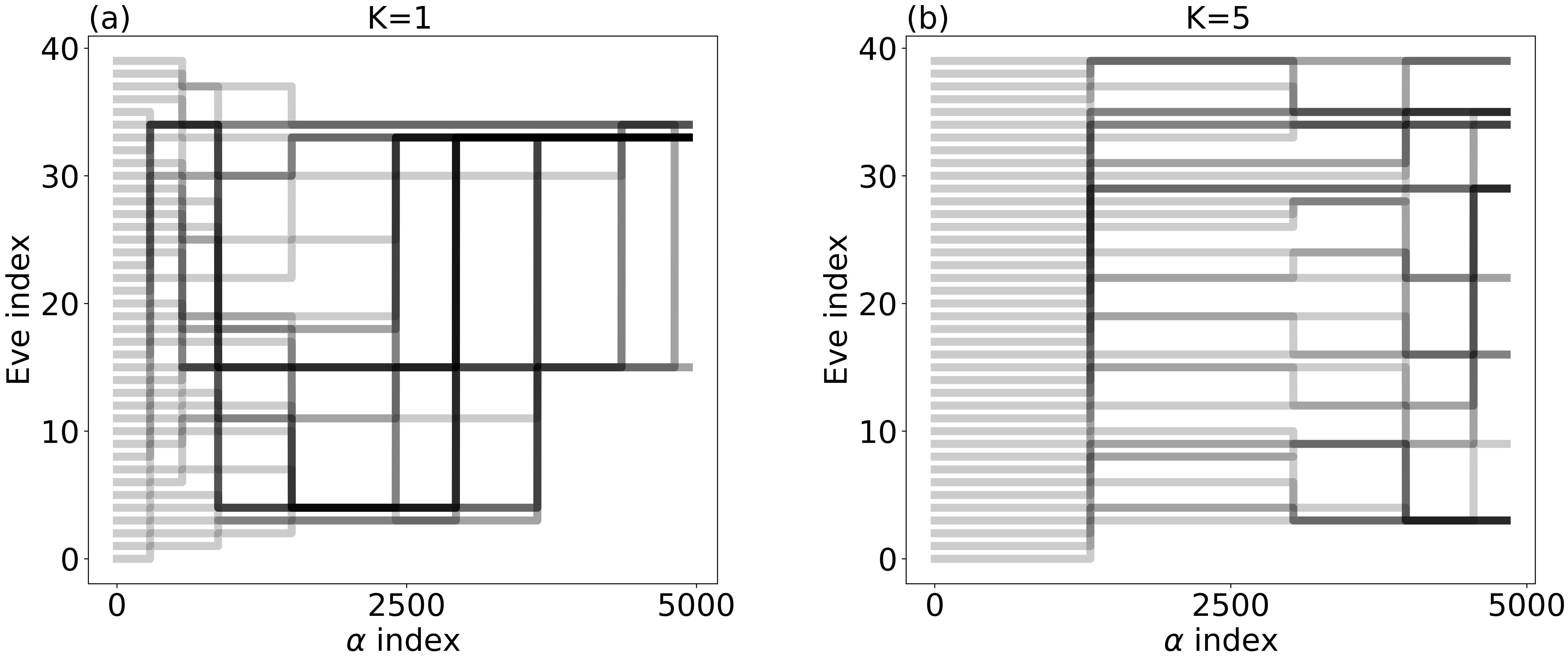}
    \caption{Eve index evolution vs. $\alpha$-sequence evolution for (a) $K=1$ and (b) $K=5$. The increasing opacity indicates superposition, that is, replication of specific Eve indices for different particles.}
\label{eve_sup}
\end{figure*}

\begin{figure*}
    \centering 
    \includegraphics[scale=0.90]{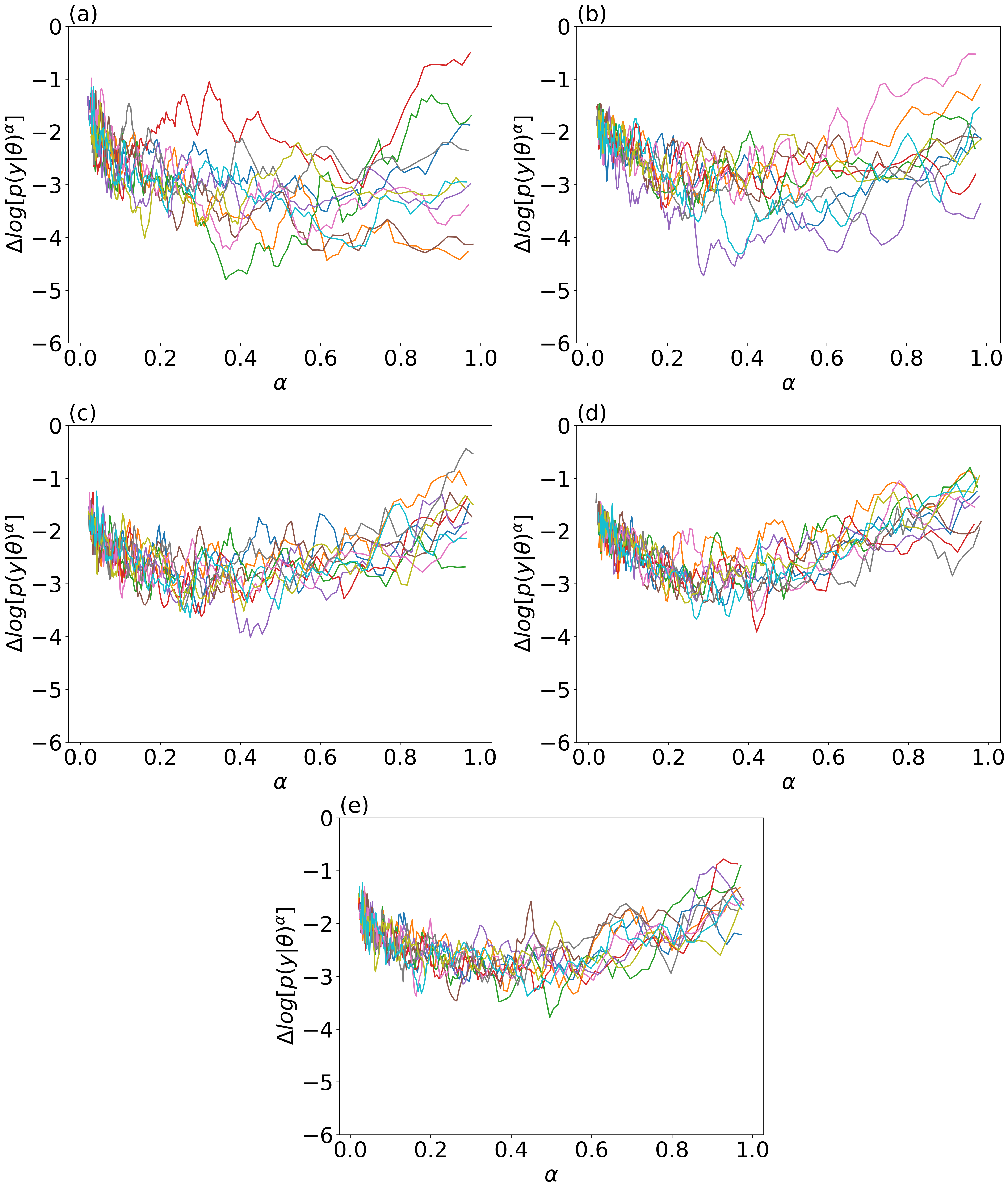}
    \caption{Evolution of the difference between the weighted mean log-likelihood $\hat p(\mathbf{y}|\mathbf{{\theta}})$ and the target log-likelihood calculated with the noise realization $p_{n}(\mathbf{y}|\mathbf{{\theta}})$ raised to $\alpha$, where each color represents one replication, for (a) $K=1$, (b) $K=3$, (c) $K=5$ (d) $K=10$, and (e) $K=20$}
\label{diff_k}
\end{figure*}

From a computational standpoint, it is beneficial if high-quality uncertainty estimates of the evidences would be obtained from one ASMC run only. Hence, we assess how the predictions of equations \ref{var} and \ref{sigma} compare with the estimates based on Monte Carlo replications.
For smaller $K$, resampling compensates for the faster increasing variance of the weights, but this is at the expense of strong correlations between the particles. The impact of resampling on the variance estimation in equation \ref{var} is primarily embodied in the sum involving the Eve indices. For smaller $K$, more resampling is needed and the number of remaining Eve indices are smaller. Figure \ref{eve_sup} illustrates the evolution of the Eve indices $E_t^{i}$ for $K=1$ and $K=5$ as the CM1 $\alpha$-sequence progresses. Of the original 40 Eve indices, there are at the end only 3 and 8 Eve indices surviving for $K=1$ and $K=5$, respectively. For $K=20$, there are $15$ surviving Eve indeces. The larger the number of surviving Eve indices, the less is the risk of mode collapse in which the ASMC algorithm only explore a small part of the posterior distribution. This basically implies that the higher-quality estimates are obtained by using larger $K$ or $CESS_{op}$, but this comes at the cost of an increasing number of forward simulations.  
Table \ref{evid_err} shows the relative standard deviation obtained with Monte Carlo replication and the single ASMC run estimates. For CM1, the relative standard deviations calculated with both estimators are similar for $K=10$ and $K=20$ suggesting that equations \ref{var} and \ref{sigma} may provide high-quality uncertainty estimates for long-enough ASMC runs. For small $K$, we observe significant underestimation of the relative standard deviations. For $K=1$, the single ASMC estimation is three times smaller than those obtained by Monte Carlo replication. 
Why does the single-run ASMC uncertainty estimation work well for large $K$, but not for small ones? To shed some light on this question, we present in Figure \ref{diff_k} the evolution of the difference between the weighted mean of the $40$ particles' likelihoods $\hat p(\mathbf{y}|\mathbf{{\theta}})$ and the target log-likelihood calculated with the noise realization $p_{n}(\mathbf{y}|\mathbf{{\theta}})$, both raised to the power of the corresponding ${\alpha}$ with the differences expressed in logarithmic units, that is, $\Delta log[p(\mathbf{y}|\mathbf{{\theta}})^\alpha] =log[\hat p(\mathbf{y}|\mathbf{{\theta}})^\alpha] - log[p_{n}(\mathbf{y}|\mathbf{{\theta}})^{\alpha}] $. This difference is shown for the ten replications and for the different $K$-values considered. In addition, Table \ref{evid_err} shows the variance and the root-mean-square error (RMSE) for the last states ($\alpha=1$) $\Delta log[p(\mathbf{y}|\mathbf{{\theta}}])$ that decrease with increasing $K$. We observe in Figure \ref{diff_k} that when $K$ decreases, the trajectories becomes more separate and show more auto-correlation. At $K=20$ and $K=10$ for which the single-ASMC estimates worked well, we observe that the trajectories overlap and cross each other, thereby, suggesting that the information content of one individual ASMC run is not so much different than another. In contrast, for $K=1$ (Figure \ref{diff_k}a) the mean trajectories tend to be more separated from each other suggesting that they sample slightly different posteriors. The Monte Carlo replications account for these differences between individual ASMC runs, while this is impossible when considering estimates from a single ASMC run. This suggests then that the single-run evidence estimator should only be trusted when performing a sufficient number of $K$ iterations, thereby, ensuring that the approximations of the intermediate distributions for different ASMC runs are small. In practice, this suggests that it is useful to run at least two ASMC runs and to ensure that the weighted mean-likelihoods of their particles are similar and tend to cross multiple times during the ASMC runs. If this is not the case, our results suggest that the uncertainty estimation of the evidence obtained from one ASMC run is too small.

This finding is also supported by the CM2 estimations in Table \ref{evid_err}. This is clearly a more challenging conceptual model, where the $K$ used for the ASMC runs was three times higher than for CM1. Even if the single-run uncertainty estimations decrease consistently when increasing $K$, the values are too low compared to those of Monte Carlo replication. This suggests that $K$ was not large enough to trust the single-run estimator. This is also reflected in the higher variance and the RMSE of the likelihood difference compared to CM1. This suggests that either Monte Carlo replications are needed to obtain an accurate error estimator or $K$ should be increased to improve the reliability of the single-run estimator.

\begin{table*}
%\begin{minipage}{106mm}
\caption{Relative standard deviation of evidence estimations obtained with ASMC-DREAM using different $K$ iterations per intermediate distribution. Results are shown for estimates based on a single run (equations \ref{sigma} and \ref{var}) and by ten replications for CM1 and five replications for CM2 of the ASMC algorithm. Variance and root-mean-square error (RMSE) of the difference between the average log-likelihoods and the target (noise) log-likelihood are shown for the replications.}
\label{}
\centering
%\resizebox{\columnwidth}{!}
\begin{tabular}{@{}ccccc}
\hline
$K$ & $\sigma_r$ [single run] &  $\sigma_r$ [replications]  & $\sigma^2 (\Delta log[p(y|\mathbf{{\theta}})])$ & $RMSE (\Delta log[p(y|\mathbf{{\theta}})])$   \\\hline  \multicolumn{5}{c}{CM1}\\\hline
1 & 0.62 & 1.72 & 1.70 & 2.99\\[2pt] 
3 & 0.42  & 0.66 & 1.42 & 1.91\\[2pt] 
5 & 0.35  & 0.50 & 0.62 & 1.84 \\[2pt] 
10 & 0.29  & 0.27 & 0.67 & 1.14 \\[2pt] 
20 & 0.21  & 0.20 & 0.69 & 1.47 \\\hline 
 \multicolumn{5}{c}{CM2}\\\hline
5& 0.47 & 1.92 & 8.45 &43.34\\[2pt]
10& 0.40 & 1.56 &4.16 & 21.59\\[2pt]
20& 0.38 & 1.02 & 3.66 & 13.89\\[2pt]
40& 0.36 & 1.52& 5.06 & 7.70\\[2pt]
60& 0.33 & 1.22 &  6.36 & 2.46\\[2pt]
\hline
\label{evid_err}
\end{tabular}
%\end{minipage}
\end{table*}

\section{Discussion}
Our results suggest that ASMC can provide accurate approximations of posterior PDFs for challenging inverse problems for which state-of-the-art adaptive MCMC fails to converge when considering a similar number of forward simulations (Figure \ref{last_models}). Furthermore, ASMC is very well suited for parallel computation, which is less the case for most MCMC methods. A general recommendation for practical applications is that the algorithmic variables $K$ and $CESS_{op}$ in Algorithm 1 are chosen sufficiently large to ensure that the weighted-mean likelihood of the particles is close to the target likelihood during the ASMC run (Figure \ref{diff_k}). Clearly, if the total number of forward simulations are insufficient, the ASMC algorithm fails in sampling posterior realizations of high likelihood for most particles. This leads to an impoverished particle approximation of the posterior PDF as evidenced by few surviving Eve indices (Figure \ref{eve_sup}) and mode collapse.  

A similar argument holds for the evidence estimation. ASMC provides an unbiased estimation, as shown for the high-noise setting example (section \ref{evid_sect}). However, the evidence estimation procedure will only be reliable if the particles approximate the target power posteriors well enough. In addition, too low $K$ and $CESS_{op}$ lead to frequent resampling that increases the estimation variance. Our results also suggest that error approximations based on single ASMC runs (eqs. \ref{var} and \ref{sigma}) are too optimistic in such settings, but reliable for sufficiently long ASMC runs (Table \ref{evid_err}). We also note that the relative standard deviations of the evidence estimates (Figure \ref{replications}) are several orders of magnitude smaller than the evidences obtained for the consistent and inconsistent prior models (Table \ref{lik_evid}).

Providing practical recommendations for parameter settings away from easily-recognizable degenerate conditions is challenging. Of course, the larger the $N$ the better, as the particle approximation of the parameter space will be improved. Our choice of $N=40$ was dictated by the number of forward runs we could perform in parallel on one compute node, while much larger values are possible on modern computational architectures. An important point is how well the posterior can be described by a weighted average of $N$ particles. The complexity of the posterior distribution depends on several factors like the dimension of the parameter space, the physics, the number and type of data, and the experimental design. Consequently, a much larger number of particles might be needed in challenging high-dimensional settings with strong parameter correlations or for problems with multi-modal posterior PDFs. In agreement with \citet{neal2001annealed}, we recommend distributing the total number of forward runs for each ASMC particle by favouring a large number of intermediate distributions over larger $K$. In practice, we typically first choose a suitably large $CESS_{op}$ and then vary $K$. In contrast to $K$, the influence of $CESS_{op}$ on the total number of forward simulations is non-linear and difficult to predict before running the algorithm. The trial tests in this study suggest that $CESS_{op}$ needs to be larger than $0.99 N$, for our considered ranges of $K$, in order to reach the target misfit and build a smooth $\alpha$-sequence. After fixing $CESS_{op}$, one can then first run the ASMC with an initially small $K$ before re-running it with a twice as large value. If the difference between the resulting evidence estimates for these two choices of $K$ are much smaller than the computed evidences for competing conceptual models, and if the inferred posteriors are similar, then this choice of $K$ is probably sufficient. If important differences are observed between the ASMC runs obtained for the different $K$, then one needs to further double $K$, and so on. Finally, the proposal scale $\epsilon$ needs to be initialized with a high enough value such that the initial acceptance rate is above $AR_{min}$. After this, the automatic rescaling of this parameter ensures high-quality estimates regardless of the model proposal scheme. 

The observed relative insensitivity of the ASMC results to the model proposal type (Figures \ref{ASMC_DREAM} and \ref{ASMC_gauss}) is noteworthy, as the MCMC results (Figure \ref{MCMC_without_ASMC}) are highly sensitive to this choice. CM1 and CM2 present different levels of complexity. For CM1, MCMC-DREAM achieves convergence without difficulty (Fig. \ref{MCMC_without_ASMC}i), while this is far from being the case for MCMC-Gauss (Figure \ref{MCMC_without_ASMC}j). For CM2, both MCMC approaches fail (Figures \ref{MCMC_without_ASMC}k and l), while ASMC-DREAM and ASMC-Gauss perform similarly well for both CM1 and CM2 (Figures \ref{ASMC_DREAM} and \ref{ASMC_gauss}). The underlying reason for the success of ASMC and its insensitivity to the proposal mechanism is likely found due to the following factors. On the one hand, the adaptive scaling of the proposals (e.g., Figure \ref{ASMC_DREAM}c) and the tempering (e.g., Figure \ref{ASMC_DREAM}d) allow the particles to more easily move away from local minima, while resampling, on the other hand, gives priority to the high-likelihood regions (e.g., Figure \ref{ASMC_DREAM}h). Clearly, no such tuning of the proposal scale is possible when using MCMC as it violates detailed balance conditions. We stress that the comparisons made herein are with MCMC algorithms running at a unitary temperature, while parallel tempering-based MCMC methods might not have these problems \citep{sambridge2014parallel}.

The presented ASMC method share similarities to other approaches for evidence estimation. Nested Sampling \citep{skilling2004nested} reduces the evidence multidimensional integral to sampling of a one-dimensional integral over prior mass elements, using an increasing constraint on the log-likelihood lower bound. Other methods rely on MCMC sampling using power posteriors. For instance, thermodynamic integration (TIE) \citep{gelman1998simulating}, also called path sampling, reduces the evidence computation to a one-dimensional integral of the expectation of the likelihood over $\alpha$. \citet{zeng2018improved} shows that TIE performs better than nested sampling in terms of accuracy and stability. Stepping Stone Sampling (SS) \citep{xie2011improving} also rely on power-posteriors but improves in accuracy compared with TIE by formulating the evidence estimation by the product of ratios of intermediate normalizing constants, that is, similarly to AIS and ASMC. An important practical difference is that SS is often performed in parallel by running multiple MCMC runs targeting different power posteriors \citep{brunetti2019hydrogeological}. Since each chain starts from the prior, the total computational cost is high, and perhaps more importantly, there is no solution to deal with MCMC chains for $\alpha$ close to one that do not converge (as in our MCMC trials with both MCMC-Gauss and MCMC-DREAM for CM2). This latter problem can be circumvented by running the SS algorithm sequentially using a similar tempering sequence as for ASMC. However, the $\alpha$-sequence needs to be pre-defined, while ASMC allows for adaptive tuning. Even if not presented here, we stress that the improvements offered by ASMC over AIS are drastic. Despite extensive testing and tuning of AIS parameters, we were unable to match the performance of ASMC.

\section{Conclusions}

This study demonstrates that adaptive sequential Monte Carlo (ASMC) is a powerful method to approximate the posterior PDF and estimate the evidence in non-linear geophysical inverse problems. Crosshole GPR examples in which complex geological priors are parameterized through deep generative networks are used for demonstration purposes, but the method is of wide applicability. ASMC is robust with respect to the type of model proposals used and to algorithmic settings, implying a comparatively low user effort required for tuning the algorithm for a given application. ASMC is particularly useful for moderately to strongly non-linear inverse problems and for multi-modal distributions, where targeting the posterior distribution with MCMC algorithms may result in poor convergence. For the considered examples, ASMC outperforms state-of-the-art adaptive MCMC in estimating posterior PDFs. The major advantage of ASMC compared with MCMC in a Bayesian model selection context is that it provides straightforward computation of the evidence. Reliable uncertainty estimation of evidence estimates is possible from single ASMC runs, provided that they are long enough. We hope that this study will stimulate further adaptations of sequential Monte Carlo in a geophysical context, and more specifically, lead researchers to the adaptation of ASMC when confronted with challenging inference problems and model selection tasks.

\section{Acknowledgements}

This work was supported by the Swiss National Science Foundation (project number: \href{http://p3.snf.ch/project-184574}{184574}). We are grateful to Prof. Arnaud Doucet (University of Oxford) who provided highly valuable suggestions at an early stage of this research. We also thank Prof. Adam Johansen (University of Warwick) for responding to an inquiry concerning the original ASMC paper. Finally, we would like to thank the editor Juan Carlos Afonso and two anonymous referees for their constructive comments. Our ASMC code and test examples are available in the following GitHub repository: https://github.com/amaya-macarena/ASMC.

\newpage

\bibliographystyle{apalike}

\bibliography{Revised_Manuscript}

\begin{thebibliography}{}

\bibitem[Bergen et~al., 2019]{bergen2019machine}
Bergen, K.~J., Johnson, P.~A., Maarten, V., and Beroza, G.~C. (2019).
\newblock Machine learning for data-driven discovery in solid earth geoscience.
\newblock {\em Science}, 363(6433).

\bibitem[Brown and Neal, 1991]{Brown}
Brown, D. and Neal, A. (1991).
\newblock The analysis of the variance and covariance of products.
\newblock {\em Biometrics}, 47(2):429--444.

\bibitem[Brunetti et~al., 2019]{brunetti2019hydrogeological}
Brunetti, C., Bianchi, M., Pirot, G., and Linde, N. (2019).
\newblock Hydrogeological model selection among complex spatial priors.
\newblock {\em Water Resources Research}, 55(8):6729--6753.

\bibitem[Brunetti et~al., 2017]{brunetti2017bayesian}
Brunetti, C., Linde, N., and Vrugt, J.~A. (2017).
\newblock Bayesian model selection in hydrogeophysics: Application to
  conceptual subsurface models of the {S}outh {O}yster {B}acterial {T}ransport
  {S}ite, {V}irginia, {USA}.
\newblock {\em Advances in Water Resources}, 102:127--141.

\bibitem[Chan and Lai, 2013]{chan2013}
Chan, H.~P. and Lai, T.~L. (2013).
\newblock A general theory of particle filters in hidden markov models and some
  applications.
\newblock {\em Ann. Statist.}, 41(6):2877--2904.

\bibitem[Curtis and Lomax, 2001]{curtis2001prior}
Curtis, A. and Lomax, A. (2001).
\newblock Prior information, sampling distributions, and the curse of
  dimensionality.
\newblock {\em Geophysics}, 66(2):372--378.

\bibitem[Del~Moral et~al., 2006]{del2006sequential}
Del~Moral, P., Doucet, A., and Jasra, A. (2006).
\newblock Sequential {M}onte {C}arlo samplers.
\newblock {\em Journal of the Royal Statistical Society: Series B (Statistical
  Methodology)}, 68(3):411--436.

\bibitem[{Douc} and {Cappe}, 2005]{Douc}
{Douc}, R. and {Cappe}, O. (2005).
\newblock Comparison of resampling schemes for particle filtering.
\newblock In {\em ISPA 2005. Proceedings of the 4th International Symposium on
  Image and Signal Processing and Analysis, 2005.}, pages 64--69.

\bibitem[Doucet and Johansen, 2011]{doucet2011tutorial}
Doucet, A. and Johansen, A.~M. (2011).
\newblock A tutorial on particle filtering and smoothing: Fifteen years later.
\newblock {\em The Oxford Handbook of Nonlinear Filtering}, 12(656-704):3.

\bibitem[Doucet and Lee, 2018]{doucet2018sequential}
Doucet, A. and Lee, A. (2018).
\newblock Sequential {M}onte {C}arlo methods.
\newblock {\em Handbook of Graphical Models}, pages 165--189.

\bibitem[Earl and Deem, 2005]{earl2005parallel}
Earl, D.~J. and Deem, M.~W. (2005).
\newblock Parallel tempering: Theory, applications, and new perspectives.
\newblock {\em Physical Chemistry Chemical Physics}, 7(23):3910--3916.

\bibitem[Gelman and Meng, 1998]{gelman1998simulating}
Gelman, A. and Meng, X.-L. (1998).
\newblock Simulating normalizing constants: From importance sampling to bridge
  sampling to path sampling.
\newblock {\em Statistical science}, pages 163--185.

\bibitem[Gelman and Rubin, 1992]{gelman1992inference}
Gelman, A. and Rubin, D.~B. (1992).
\newblock Inference from iterative simulation using multiple sequences.
\newblock {\em Statistical Science}, 7(4):457--472.

\bibitem[Goodfellow et~al., 2016]{Goodfellow-et-al-2016}
Goodfellow, I., Bengio, Y., and Courville, A. (2016).
\newblock {\em Deep Learning}.
\newblock MIT Press.
\newblock \url{http://www.deeplearningbook.org}.

\bibitem[Goodfellow et~al., 2014]{goodfellow2014generative}
Goodfellow, I., Pouget-Abadie, J., Mirza, M., Xu, B., Warde-Farley, D., Ozair,
  S., Courville, A., and Bengio, Y. (2014).
\newblock Generative adversarial nets.
\newblock In {\em Advances in Neural Information Processing Systems}, pages
  2672--2680.

\bibitem[Hammersley and Handscomb, 1964]{Hammersley1964}
Hammersley, J.~M. and Handscomb, D.~C. (1964).
\newblock {\em General Principles of the {M}onte {C}arlo Method}, pages 50--75.
\newblock Springer Netherlands, Dordrecht.

\bibitem[Hansen et~al., 2012]{hansen2012inverse}
Hansen, T.~M., Cordua, K.~S., and Mosegaard, K. (2012).
\newblock Inverse problems with non-trivial priors: efficient solution through
  sequential {G}ibbs sampling.
\newblock {\em Computational Geosciences}, 16(3):593--611.

\bibitem[Jetchev et~al., 2016]{jetchev2016texture}
Jetchev, N., Bergmann, U., and Vollgraf, R. (2016).
\newblock Texture synthesis with spatial generative adversarial networks.
\newblock {\em arXiv preprint arXiv:1611.08207}.

\bibitem[Karpatne et~al., 2018]{karpatne2018machine}
Karpatne, A., Ebert-Uphoff, I., Ravela, S., Babaie, H.~A., and Kumar, V.
  (2018).
\newblock Machine learning for the geosciences: Challenges and opportunities.
\newblock {\em IEEE Transactions on Knowledge and Data Engineering},
  31(8):1544--1554.

\bibitem[Kingma and Welling, 2013]{kingma2013auto}
Kingma, D.~P. and Welling, M. (2013).
\newblock Auto-encoding variational {B}ayes.
\newblock {\em arXiv preprint arXiv:1312.6114}.

\bibitem[Kirkpatrick et~al., 1983]{kirkpatrick1983optimization}
Kirkpatrick, S., Gelatt, C.~D., and Vecchi, M.~P. (1983).
\newblock Optimization by simulated annealing.
\newblock {\em Science}, 220(4598):671--680.

\bibitem[Koltermann and Gorelick, 1996]{koltermann1996heterogeneity}
Koltermann, C.~E. and Gorelick, S.~M. (1996).
\newblock Heterogeneity in sedimentary deposits: A review of
  structure-imitating, process-imitating, and descriptive approaches.
\newblock {\em Water Resources Research}, 32(9):2617--2658.

\bibitem[Kong et~al., 1994]{kong1994sequential}
Kong, A., Liu, J.~S., and Wong, W.~H. (1994).
\newblock Sequential imputations and {B}ayesian missing data problems.
\newblock {\em Journal of the American Statistical Association},
  89(425):278--288.

\bibitem[Laloy et~al., 2018]{laloy2018training}
Laloy, E., H{\'e}rault, R., Jacques, D., and Linde, N. (2018).
\newblock Training-image based geostatistical inversion using a spatial
  generative adversarial neural network.
\newblock {\em Water Resources Research}, 54(1):381--406.

\bibitem[Laloy et~al., 2017]{laloy2017inversion}
Laloy, E., H{\'e}rault, R., Lee, J., Jacques, D., and Linde, N. (2017).
\newblock Inversion using a new low-dimensional representation of complex
  binary geological media based on a deep neural network.
\newblock {\em Advances in Water Resources}, 110:387--405.

\bibitem[Laloy et~al., 2019]{laloy2019gradient}
Laloy, E., Linde, N., Ruffino, C., H{\'e}rault, R., Gasso, G., and Jacques, D.
  (2019).
\newblock Gradient-based deterministic inversion of geophysical data with
  generative adversarial networks: Is it feasible?
\newblock {\em Computers \& Geosciences}, 133:104333.

\bibitem[Laloy and Vrugt, 2012]{laloy2012high}
Laloy, E. and Vrugt, J.~A. (2012).
\newblock High-dimensional posterior exploration of hydrologic models using
  multiple-try {DREAM(ZS)} and high-performance computing.
\newblock {\em Water Resources Research}, 48(1).

\bibitem[LeCun et~al., 2015]{lecun2015deep}
LeCun, Y., Bengio, Y., and Hinton, G. (2015).
\newblock Deep learning.
\newblock {\em nature}, 521(7553):436--444.

\bibitem[Lee and Whiteley, 2018]{lee2018variance}
Lee, A. and Whiteley, N. (2018).
\newblock Variance estimation in the particle filter.
\newblock {\em Biometrika}, 105(3):609--625.

\bibitem[Lewis and Raftery, 1997]{lewis1997estimating}
Lewis, S.~M. and Raftery, A.~E. (1997).
\newblock Estimating {B}ayes factors via posterior simulation with the
  {L}aplace—{M}etropolis estimator.
\newblock {\em Journal of the American Statistical Association},
  92(438):648--655.

\bibitem[Linde et~al., 2017]{linde2017uncertainty}
Linde, N., Ginsbourger, D., Irving, J., Nobile, F., and Doucet, A. (2017).
\newblock On uncertainty quantification in hydrogeology and hydrogeophysics.
\newblock {\em Advances in Water Resources}, 110:166--181.

\bibitem[Linde et~al., 2015]{linde2015geological}
Linde, N., Renard, P., Mukerji, T., and Caers, J. (2015).
\newblock Geological realism in hydrogeological and geophysical inverse
  modeling: A review.
\newblock {\em Advances in Water Resources}, 86:86--101.

\bibitem[Mariethoz and Caers, 2014]{mariethoz2014multiple}
Mariethoz, G. and Caers, J. (2014).
\newblock {\em Multiple-point geostatistics: Stochastic modeling with training
  images}.
\newblock John Wiley \& Sons.

\bibitem[Mariethoz et~al., 2010]{mariethoz2010bayesian}
Mariethoz, G., Renard, P., and Caers, J. (2010).
\newblock Bayesian inverse problem and optimization with iterative spatial
  resampling.
\newblock {\em Water Resources Research}, 46(11).

\bibitem[Mosegaard and Tarantola, 1995]{mosegaard1995monte}
Mosegaard, K. and Tarantola, A. (1995).
\newblock Monte carlo sampling of solutions to inverse problems.
\newblock {\em Journal of Geophysical Research: Solid Earth},
  100(B7):12431--12447.

\bibitem[Mosser et~al., 2017]{mosser2017reconstruction}
Mosser, L., Dubrule, O., and Blunt, M.~J. (2017).
\newblock Reconstruction of three-dimensional porous media using generative
  adversarial neural networks.
\newblock {\em Physical Review E}, 96(4):043309.

\bibitem[Mosser et~al., 2020]{mosser2020stochastic}
Mosser, L., Dubrule, O., and Blunt, M.~J. (2020).
\newblock Stochastic seismic waveform inversion using generative adversarial
  networks as a geological prior.
\newblock {\em Mathematical Geosciences}, 52(1):53--79.

\bibitem[Neal, 2001]{neal2001annealed}
Neal, R.~M. (2001).
\newblock Annealed importance sampling.
\newblock {\em Statistics and Computing}, 11(2):125--139.

\bibitem[Peterson, 2001]{peterson2001pre}
Peterson, Jr, J.~E. (2001).
\newblock Pre-inversion corrections and analysis of radar tomographic data.
\newblock {\em Journal of Environmental \& Engineering Geophysics}, 6(1):1--18.

\bibitem[Pirot et~al., 2019]{pirot2019reduction}
Pirot, G., Huber, E., Irving, J., and Linde, N. (2019).
\newblock Reduction of conceptual model uncertainty using ground-penetrating
  radar profiles: Field-demonstration for a braided-river aquifer.
\newblock {\em Journal of Hydrology}, 571:254--264.

\bibitem[Pirot et~al., 2015]{pirot2015pseudo}
Pirot, G., Straubhaar, J., and Renard, P. (2015).
\newblock A pseudo genetic model of coarse braided-river deposits.
\newblock {\em Water Resources Research}, 51(12):9595--9611.

\bibitem[Podvin and Lecomte, 1991]{podvin1991finite}
Podvin, P. and Lecomte, I. (1991).
\newblock Finite difference computation of traveltimes in very contrasted
  velocity models: a massively parallel approach and its associated tools.
\newblock {\em Geophysical Journal International}, 105(1):271--284.

\bibitem[Renard and Allard, 2013]{renard2013connectivity}
Renard, P. and Allard, D. (2013).
\newblock Connectivity metrics for subsurface flow and transport.
\newblock {\em Advances in Water Resources}, 51:168--196.

\bibitem[Roth et~al., 1990]{roth1990calibration}
Roth, K., Schulin, R., Fl{\"u}hler, H., and Attinger, W. (1990).
\newblock Calibration of time domain reflectometry for water content
  measurement using a composite dielectric approach.
\newblock {\em Water Resources Research}, 26(10):2267--2273.

\bibitem[Sambridge, 2014]{sambridge2014parallel}
Sambridge, M. (2014).
\newblock A parallel tempering algorithm for probabilistic sampling and
  multimodal optimization.
\newblock {\em Geophysical Journal International}, 196(1):357--374.

\bibitem[Sch{\"o}niger et~al., 2014]{schoniger2014model}
Sch{\"o}niger, A., W{\"o}hling, T., Samaniego, L., and Nowak, W. (2014).
\newblock Model selection on solid ground: Rigorous comparison of nine ways to
  evaluate {B}ayesian model evidence.
\newblock {\em Water Resources Research}, 50(12):9484--9513.

\bibitem[Scott, 2015]{scott2015multivariate}
Scott, D.~W. (2015).
\newblock {\em Multivariate Density Estimation: Theory, Practice, and
  Visualization}.
\newblock John Wiley \& Sons.

\bibitem[Skilling, 2004]{skilling2004nested}
Skilling, J. (2004).
\newblock Nested sampling.
\newblock In {\em AIP Conference Proceedings}, volume 735, pages 395--405.
  American Institute of Physics.

\bibitem[Vrugt, 2016]{vrugt2016markov}
Vrugt, J.~A. (2016).
\newblock Markov chain {Monte} {Carlo} simulation using the {DREAM} software
  package: Theory, concepts, and {MATLAB} implementation.
\newblock {\em Environmental Modelling \& Software}, 75:273--316.

\bibitem[Xie et~al., 2011]{xie2011improving}
Xie, W., Lewis, P.~O., Fan, Y., Kuo, L., and Chen, M.-H. (2011).
\newblock Improving marginal likelihood estimation for {B}ayesian phylogenetic
  model selection.
\newblock {\em Systematic Biology}, 60(2):150--160.

\bibitem[Zahner et~al., 2016]{zahner2016image}
Zahner, T., Lochb{\"u}hler, T., Mariethoz, G., and Linde, N. (2016).
\newblock Image synthesis with graph cuts: a fast model proposal mechanism in
  probabilistic inversion.
\newblock {\em Geophysical Journal International}, 204(2):1179--1190.

\bibitem[Zeng et~al., 2018]{zeng2018improved}
Zeng, X., Ye, M., Wu, J., Wang, D., and Zhu, X. (2018).
\newblock Improved nested sampling and surrogate-enabled comparison with other
  marginal likelihood estimators.
\newblock {\em Water Resources Research}, 54(2):797--826.

\bibitem[Zhou et~al., 2016]{zhou2016toward}
Zhou, Y., Johansen, A.~M., and Aston, J.~A. (2016).
\newblock Toward automatic model comparison: an adaptive sequential {M}onte
  {C}arlo approach.
\newblock {\em Journal of Computational and Graphical Statistics},
  25(3):701--726.

\end{thebibliography}

\end{document}